\ifdefined\JCPPreprint
\documentclass[aip,jcp,preprint,amsmath,amssymb,floatfix]{revtex4-2}
\else
\documentclass[aip,jcp,reprint,amsmath,amssymb,floatfix]{revtex4-2}
\fi
\usepackage[T1]{fontenc}
\usepackage[utf8]{inputenc}
\usepackage{mathrsfs,amsthm,mathtools,bm,booktabs,graphicx,microtype,xurl}
\usepackage[hidelinks]{hyperref}
\newtheorem{theorem}{Theorem}
\newtheorem{proposition}{Proposition}

\DeclareMathOperator{\conv}{conv}
\DeclareMathOperator{\diag}{diag}
\DeclareMathOperator{\Tr}{Tr}
\DeclareMathOperator{\Var}{Var}
\DeclareMathOperator{\Sym}{Sym}

\newcommand{\R}{\mathbb R}
\newcommand{\Z}{\mathbb Z}
\newcommand{\E}{\mathbb E}
\newcommand{\1}{\mathbf1}
\DeclareMathAlphabet{\mathpzc}{OT1}{pzc}{m}{it}
\newcommand{\cfg}{\mathpzc n}
\newcommand{\dd}{\,\mathrm d}
\newcommand{\Lcal}{\mathcal L}

\newcommand{\CN}[1]{\mathcal C_{N,#1}}

\newcommand{\gfun}[1]{\gamma[#1]}
\newcommand{\dop}[1]{\widehat d(#1)}
\hypersetup{pdftitle={Integer electron-count cuts in variational second-order reduced-density-matrix theory. II. A convergent hierarchy of rigorous relaxations for real-space pair-density representability},pdfauthor={Maho Nakata}}
\newcommand{\adaptCutsnopPQG}{6}

\newcommand{\adaptDeltaTwonop}{4.2\times10^{-4}}

\newcommand{\adaptPlaneOnenop}{3.93}
\newcommand{\adaptPlaneTwonop}{3.00}

\newcommand{\adaptStepBoundnop}{10^{-4}}

\newcommand{\adaptStepnop}{0.2}

\newcommand{\cmpAdaptGapnopThreeZero}{0.07}

\newcommand{\cmpAdaptcnmFourZero}{-0.084}

\newcommand{\cmpAdaptnTwoFourZero}{-22.05}
\newcommand{\cmpAdaptnopTwoFive}{-8.930}
\newcommand{\cmpAdaptnopThreeZero}{-0.979}
\newcommand{\cmpAdaptnopThreeFive}{-0.094}
\newcommand{\cmpAdaptnopFourZero}{-0.014}

\newcommand{\cmpBasebeoFourZero}{-32.03}

\newcommand{\cmpFixedcnmFourZero}{-0.333}

\newcommand{\cmpFixedcoInf}{-19.51}
\newcommand{\cmpFixednTwoFourZero}{-22.05}

\newcommand{\cmpFixednopThreeZero}{-2.103}

\newcommand{\cmpFixednopInf}{-0.001}
\newcommand{\cmpPartITtbeoFourZero}{-0.598}
\newcommand{\cmpPartITtcnmFourZero}{-0.145}

\newcommand{\cmpPartITtnopTwoFive}{-11.06}
\newcommand{\cmpPartITtnopThreeZero}{-2.685}
\newcommand{\cmpPartITtnopThreeFive}{-0.560}
\newcommand{\cmpPartITtnopFourZero}{-0.124}

\newcommand{\cmpPartIcnmFourZero}{-0.218}

\newcommand{\cmpPartInopTwoFive}{-15.76}
\newcommand{\cmpPartInopThreeZero}{-3.496}
\newcommand{\cmpPartInopThreeFive}{-0.683}
\newcommand{\cmpPartInopFourZero}{-0.146}

\newcommand{\cmpTtBasebeoFourZero}{-17.51}

\newcommand{\cmpTtbeoFourZero}{-0.153}
\newcommand{\cmpTtcnmFourZero}{-0.032}

\newcommand{\cmpTtcoInf}{-5.013}
\newcommand{\cmpTtnTwoFourZero}{-1.625}
\newcommand{\cmpTtnTwoInf}{-1.433}
\newcommand{\cmpTtnopTwoFive}{-5.139}
\newcommand{\cmpTtnopThreeZero}{-0.458}
\newcommand{\cmpTtnopThreeFive}{-0.035}
\newcommand{\cmpTtnopFourZero}{-0.003}

\newcommand{\cmpTwoCellnopTwoFive}{-8.955}
\newcommand{\cmpTwoCellnopThreeZero}{-0.990}

\newcommand{\cutFigExcessCuts}{0.009}
\newcommand{\cutFigExcessPqg}{0.397}

\newcommand{\ePdeltaPqgnTwoFourZero}{0.9956}

\newcommand{\ePfineCellsnTwoFourZero}{896}

\newcommand{\ePfinenTwoFourZero}{3.2}

\newcommand{\ePmassnTwoFourZero}{156}

\newcommand{\ePslabCutsbeoFourZero}{9.5}
\newcommand{\ePslabCutsnTwoFourZero}{13.0}
\newcommand{\ePslabCutsnopFourZero}{13.3}

\newcommand{\ePslabnTwoFourZero}{12.8}

\newcommand{\fciDistCutsbeoFourZero}{0.224}
\newcommand{\fciDistCutsnTwoFourZero}{14.6}
\newcommand{\fciDistCutsnopFourZero}{0.019}

\newcommand{\fciDistPqgnopFourZero}{16.5}

\newcommand{\fixRhoBoundCutsbeoFourZero}{0.458}

\newcommand{\fixRhoBoundCutsnopFourZero}{0.025}

\newcommand{\fracBcutMax}{29.2}
\newcommand{\fracBcutMin}{2.1}
\newcommand{\fracBcutOwnMax}{24.1}
\newcommand{\fracBcutOwnMin}{2.0}
\newcommand{\fracBcutTMax}{19.3}
\newcommand{\fracBcutTMin}{0.3}

\newcommand{\fracBfockMax}{73.6}
\newcommand{\fracBfockMin}{16.4}
\newcommand{\fracBsplitMax}{15.6}
\newcommand{\fracBsplitMin}{0.2}
\newcommand{\fracBsplitTMax}{9.4}
\newcommand{\fracBsplitTMin}{0.1}
\newcommand{\fracEndGapMax}{7.5}
\newcommand{\fracEndGapTMax}{9.2}
\newcommand{\fracZerovarMax}{71.5}
\newcommand{\fracZerovarMin}{30.5}
\newcommand{\fsBcutAbsDiff}{18.6}

\newcommand{\fsBcutMolErr}{-45.9}
\newcommand{\fsBcutMolNA}{6.709}
\newcommand{\fsBcutMolVar}{0.206}
\newcommand{\fsBcutNA}{6.625}
\newcommand{\fsStoMaxDiff}{2.3}
\newcommand{\fsZerovarAbsDiff}{3.1}

\newcommand{\fsZerovarMolNA}{6.523}
\newcommand{\fsZerovarMolVar}{0.005}
\newcommand{\fsZerovarNA}{6.500}

\newcommand{\gapNOpInf}{0.004}

\newcommand{\polyAtomsHsixpTwoFive}{-47.71}

\newcommand{\polyHalfspaceHsixpTwoFive}{-28.30}

\newcommand{\polyJointHsixpTwoFive}{-21.41}

\newcommand{\polySubsetsHsixpTwoFive}{-21.41}

\newcommand{\rankTwoCutShiftMax}{0.015}

\newcommand{\rankTwoShiftMax}{0.1}

\newcommand{\refDeltaSepFineNOp}{4.8\times10^{-5}}

\newcommand{\rvCutsGapnTwoFourZero}{5\times10^{-8}}
\newcommand{\rvCutsnTwoFourZero}{0.250000}
\newcommand{\rvFcinTwoFourZero}{0.2502}
\newcommand{\rvPqgnTwoFourZero}{0.0011}

\newcommand{\zsplitnop}{3.93}

\begin{document}
\title{Integer electron-count cuts in variational second-order reduced-density-matrix theory. II. A convergent hierarchy of rigorous relaxations for real-space pair-density representability}
\author{Maho Nakata}
\email{maho@riken.jp}
\affiliation{RIKEN Cluster for Pioneering Research, Wako, Saitama 351-0198, Japan}
\date{\today}
\begin{abstract}
Variational second-order reduced-density-matrix calculations can dissociate molecules into fractionally charged fragments; Part I largely removed this failure with integer electron-count conditions on fixed partitions. Building on the realizability theory of point processes, we place these conditions in a convergent hierarchy of rigorous relaxations, defined by finite occupation linear programs (LPs) on spatial partitions, for real-space pair densities at fixed one-particle density. Every level contains all representable pair densities, the levels tighten under nested refinement, and their intersection characterizes pair densities admitting a symmetric $N$-particle probability-measure extension. In a fixed orbital basis the levels give monotonically increasing lower bounds to the full configuration-interaction energy, which they need not reach, even for the Coulomb repulsion. The spin-summed real-space cuts of Part I are its two-cell members. The LP optimum is also a measure: at fixed one-particle density it equals the $L^1$ distance of the coarse-grained candidate from representability and increases under refinement to that of the candidate, with a vanishing coarse-graining error; its primal law constructs a representable approximation, and its dual a violated valid inequality, used as a cut in molecular calculations. The count conditions are not implied by the usual positivity conditions: moment data that satisfy all three-particle positivity conditions, including $T_1$ and $T_2$, give through Gaussian orbitals a smooth pair density that violates an integer count. For atoms at fractional electron number, ensemble relaxations with the integer-count condition lie closer to the exact piecewise-linear energy than the zero-variance fixed-$N$ form.
\end{abstract}
\maketitle

\section{Introduction}
\label{sec:intro}
Variational optimization of the second-order reduced density matrix (2-RDM) under necessary $N$-representability conditions\cite{Coleman1963} gives lower bounds to the ground-state energy.\cite{Garrod1964,GarrodFusco1976,Erdahl1978} Semidefinite programming made it a practical method,\cite{Nakata2001,Zhao2004,Mazziotti2004} first-order algorithms extended it to large systems,\cite{Mazziotti2011} and Ref.~\citenum{MazziottiBook2007} reviews the field. Van Aggelen, Bultinck, Ayers, and co-workers, however, showed that with the two-index conditions $P$, $Q$, and $G$ it dissociates diatomic molecules into fractionally charged atoms with too low an energy.\cite{VanAggelen2009,VanAggelen2010} Subsystem constraints built from precomputed fragment energies cure this for diatomic molecules,\cite{Verstichel2010,VanAggelen2011} and a recent tutorial review still lists the failure as a difficulty of the method, because such constraints are hard to apply to general polyatomic systems.\cite{DePrince2024} In Part I,\cite{NakataPartI} the integer electron-count inequality $\langle(\hat N_\Omega-k)(\hat N_\Omega-k-1)\rangle\ge0$, a long-known consequence of the integer spectrum of a number operator,\cite{Yamada1961,Erdahl1987} was imposed on fixed partitions of diatomic molecules; a single region bounded at the bond critical point removed most of the dissociation error.

Here we place these inequalities as the simplest members of a convergent hierarchy of rigorous relaxations for the real-space pair density at fixed one-particle density $\rho$, defined by finite occupation linear programs (LPs) on spatial partitions. Every level contains all representable pair densities and therefore, added to a 2-RDM relaxation in a finite basis, yields lower bounds to the full configuration-interaction energy in that basis; the levels tighten under nested refinement, and their intersection is the representable set (Proposition~\ref{prop:hierarchy}). These facts follow from known realizability and exchangeability results;\cite{Kuna2007,Kuna2011,Friesecke2013,KonstantopoulosYuan2019,Carlier2023} we establish three properties. (i) The count conditions of Part I are the spin-summed two-cell members; for three or more cells the joint conditions contain inequalities that no single-region count implies, and at fixed $S_z$ the joint law of both spins detects violations that the charge and single-spin counts miss. (ii) The LP optimum measures, and does not only detect, the distance from representability: a primal occupation law constructs a representable $N$-body density with the same $\rho$, a dual solution gives a violated valid inequality, and $\delta_{\mathcal P}\le d_\rho(n_2)\le\delta_{\mathcal P}+e_{\mathcal P}$ with $e_{\mathcal P}\to0$ under refinement (Theorem~\ref{thm:distance}). Unlike the distance to an outer approximation used to purify measured 2-RDMs,\cite{Smart2022} this bound is closed from above, and it gives a criterion for choosing partitions, which Part I chose by chemical reasoning. (iii) The count conditions are not implied by 3-positivity, the third level of the positivity hierarchy of Erdahl and Jin,\cite{ErdahlJin2000,MazziottiErdahl2001} which includes the conditions $T_1$, $T_2$, and $T_2'$ implemented by Zhao and co-workers:\cite{Zhao2004} a two-center Gaussian construction gives smooth pair densities that satisfy 3-positivity while a region has mean population $5/2$ and variance below $1/4$, the real-space signature of fractionally charged fragments.\cite{VanAggelen2009,Verstichel2010,NakataPartI} A minimal number of cuts or an optimal partition is not determined.

Each ingredient was available before Part I: regional electron-number quantization was noted for model systems,\cite{Nakata2006,VanAggelen2009} and the integer-variance bound is classical for point processes.\cite{Yamada1961,Kuna2007} We suggest that it was not used as a variational constraint because 2-RDM theory is built from positivity conditions in an orbital basis, which do not produce the count conditions [property (iii)] and in which, after projection onto a finite orbital space, a real-space count operator generally loses its integer spectrum (Sec.~\ref{sec:gaussianinputs}), and because the dissociation failure was diagnosed through energies and mean populations\cite{VanAggelen2009,VanAggelen2011} rather than through the count variance of a fragment. Real-space necessary conditions were derived by Davidson\cite{Davidson1995} and by Ayers and Davidson, whose configuration-potential family [Eqs.~(59) and (60) of Ref.~\citenum{AyersDavidson2006}] contains the dual LP as a finite, cellwise case; Friesecke and co-workers used two-set partitions and half spaces to detect a smooth nonrepresentable pair density,\cite{Friesecke2013} and Pistol constructed localized representable pair densities\cite{Pistol2004} without fixing $\rho$ or measuring a distance. In orbital space the diagonal occupations are characterized by the Slater hull,\cite{Kummer1967,YoseloffKuhn1969} an NP-hard section of the Boolean quadric polytope\cite{DezaLaurent1994a,DezaLaurent1994b,DezaLaurent1997} with explicit inequality families,\cite{WeinholdWilson1967,McRaeDavidson1972,Davidson2003,AyersDavidson2007} and complete conditions for the orbital 2-RDM, given by Erdahl,\cite{Erdahl1978,Zhao2004} were restated by Mazziotti as $(2,q)$-positivity\cite{Mazziotti2012,Mazziotti2023} and by Rubin and DePrince as sum-of-squares witnesses;\cite{RubinDePrince2026} the general problem is QMA-complete.\cite{Liu2007} The present question is the classical one for the real-space pair density, which also concerns direct pair-density models.\cite{Higuchi2007,Higuchi2017,Pederson2022,Dusson2025}

Section~\ref{sec:definitions} defines occupation laws and the finite occupation LP, shows that its two-cell members are the count conditions of Part I, and states the hierarchy (Proposition~\ref{prop:hierarchy}). Section~\ref{sec:constructive} proves the distance bracket, which closes in an exact rational example given in the supplementary material, and treats spin-resolved pair densities at fixed $S_z$. Section~\ref{sec:gaussianinputs} evaluates the cell data and cuts from orbital-basis 2-RDMs and states what the $G$ condition guarantees, and Sec.~\ref{sec:three} places a five-electron input that satisfies 3-positivity in ten Gaussian orbitals, giving a smooth pair density that violates an integer count. Section~\ref{sec:molecules} uses the LP as a separation oracle in variational 2-RDM calculations without Hartree--Fock input, where its cuts reproduce and in part improve on the region conditions of Part I, and treats atoms at fractional electron number $N=k+\theta$, where the exact energy is piecewise linear in $N$:\cite{Perdew1982,Yang2000} the fixed-$N$ conditions at fractional $N$, which as far as can be inferred underlie the analysis of Ref.~\citenum{VanAggelen2009}, give zero count variance, whereas every physical state has a variance of at least $\theta(1-\theta)$, and Fock-space relaxations that contain the physical ensembles lie considerably closer to the piecewise-linear curve. Section~\ref{sec:scope} discusses the scope of probability representability, and Sec.~\ref{sec:conclusions} concludes.

\section{Pair densities, occupation laws, and finite diagnosis}
\label{sec:definitions}
\subsection{From wavefunctions to classical electron densities}
Let $\Psi(x_1,\ldots,x_N)$ be a normalized $N$-electron wavefunction, where $x_i=(\bm r_i,\sigma_i)$ collects the position and spin of electron $i$. Its $N$-electron density $P_N=|\Psi|^2$ (for an ensemble, $\sum_ip_i|\Psi_i|^2$) is nonnegative, integrates to one, and is symmetric under any exchange of electron labels. Integrating out all but one or two electrons gives the one- and two-electron densities,
\begin{subequations}\label{eq:reduceddensities}
\begin{align}
 \rho(x)&=N\int P_N(x,x_2,\ldots)\,\dd x_2\cdots\dd x_N,\\
 n_2(x,y)&=N(N-1)\int P_N(x,y,x_3,\ldots)\,\dd x_3\cdots\dd x_N,
\end{align}
\end{subequations}
with $\int\rho=N$, $\iint n_2=M:=N(N-1)$, and $\int n_2(x,y)\,\dd y=(N-1)\rho(x)$. Here $\rho(x)\dd x$ is the mean number of electrons in $\dd x$, and $n_2(x,y)\dd x\dd y$ the mean number of ordered pairs of different electrons with one in $\dd x$ and the other in $\dd y$; they are the diagonal parts of the 1- and 2-RDMs.

Taking $|\Psi|^2$ removes the phase and the sign changes of the wavefunction, and what remains is a symmetric probability distribution for the positions of $N$ indistinguishable particles. The densities $\rho$ and $n_2$ are its one- and two-particle marginals, exactly as for $N$ classical particles; conversely, every symmetric probability density is $|\Psi|^2$ for some, possibly irregular, antisymmetric $\Psi$. Whether a given $n_2$ is the pair density of some $N$-electron state is thus a classical question: is there a symmetric probability distribution of $N$ particles whose two-particle marginal is $n_2$? We call such an $n_2$ representable, and write $\mathcal P^{\mathrm{ac}}_{N,\rho}$ for the representable pair densities with one-electron density $\rho$ whose $N$-electron distribution has a density. Finite kinetic energy and compatibility with a prescribed 1-RDM are additional requirements, discussed in Sec.~\ref{sec:scope}; distributions without a density are treated in the supplementary material.

For a candidate $n_2$, nonnegative, symmetric, and integrable with mass $M$, we fix $\rho$ to its contraction, $\rho(x)=\int n_2(x,y)\dd y/(N-1)$, and measure its distance from representability by
\begin{equation}
 d_\rho(n_2)=\inf_{\widetilde n_2\in\mathcal P^{\mathrm{ac}}_{N,\rho}}
          \|n_2-\widetilde n_2\|_{L^1(X^2)},
 \label{eq:distance_definition}
\end{equation}
the smallest total change of the ordered-pair density that makes it representable without changing $\rho$. Here $X=\R^3$ (or $\R^3$ with spin) is the one-electron space; dividing by $M$ gives probability-normalized distances.

\subsection{Distributing electrons over cells}
Now divide space into $m$ cells $A_1,\ldots,A_m$ that do not overlap and together fill all space, with characteristic functions $\chi_a$ ($\chi_a(x)=1$ in cell $a$, $0$ outside). Examples are slabs between planes, balls about nuclei, or atomic Voronoi cells; we call such a division a finite partition $\mathcal P$. In any arrangement $(x_1,\ldots,x_N)$ of the electrons, cell $a$ holds
\begin{equation}
 N_a=\sum_{i=1}^N\chi_a(x_i)
 \label{eq:cellcount}
\end{equation}
electrons, an integer between $0$ and $N$, with $\sum_{a=1}^mN_a=N$. The number $N_aN_b-\delta_{ab}N_a$ is the number of ordered pairs of different electrons with the first in cell $a$ and the second in cell $b$. Averaging gives the cell data
\begin{subequations}\label{eq:spatialmoments}
\begin{align}
 m_a&=\E[N_a]=\int_{A_a}\rho,\\
 D_{ab}&=\E[N_aN_b-\delta_{ab}N_a]=\iint_{A_a\times A_b}n_2.
\end{align}
\end{subequations}
Thus $D=D^{\mathsf T}$, every entry of $D$ is nonnegative, and $\sum_{b=1}^mD_{ab}=(N-1)m_a$. The diagonal entry $D_{aa}=\E[N_a(N_a-1)]$ counts pairs inside one cell; it is not an orbital double occupation and need not vanish. In second quantization $N_a$ is the cell number operator $\hat N_a=\int_{A_a}\hat\psi^\dagger(x)\hat\psi(x)\dd x$; the operators of different cells commute and have integer eigenvalues, and the count inequality $\langle(\hat N_\Omega-k)(\hat N_\Omega-k-1)\rangle\ge0$ of Part I expresses that $N_\Omega$ is an integer.

\subsection{Densities built from a partition}
The construction can be reversed. A configuration $\cfg=(n_1,\ldots,n_m)$ is a way of putting $N$ electrons into the cells, empty cells allowed. The number of configurations is
\begin{equation}
 K=\binom{N+m-1}{N},
 \label{eq:nconfig}
\end{equation}
the number of ways of placing $N$ electrons and $m-1$ walls between cells in a row, and we number them $\cfg^{(1)},\ldots,\cfg^{(K)}$, with $\cfg^{(j)}=(n^{(j)}_1,\ldots,n^{(j)}_m)$. Two electrons in two cells, for example, give the three configurations $(2,0)$, $(1,1)$, and $(0,2)$; three electrons in three cells give the ten configurations of Table~\ref{tab:threecell}. Choose probabilities $w_j$ for the configurations, place $n^{(j)}_a$ electrons in cell $a$ with probability $w_j$, and spread the electrons of each cell inside it according to $\rho$, independently of each other. With the normalized density of cell $a$, $\hat\rho_a(x)=\rho(x)\chi_a(x)/m_a$ (cells with $m_a=0$ may be omitted), and with the number of ordered pairs of a configuration $\cfg$ in cells $a$ and $b$,
\begin{equation}
 Q(\cfg)_{ab}=n_an_b-\delta_{ab}n_a,
 \label{eq:paircount}
\end{equation}
this gives an $N$-electron density and its one- and two-electron densities
\begin{subequations}\label{eq:densitylift}
\begin{align}
 p_{N,w}&=\sum_{j=1}^Kw_j\,\Sym\!\left(\bigotimes_{a=1}^m \hat\rho_a^{\otimes n^{(j)}_a}\right),\\
 \widetilde\rho_w(x)&=\sum_{a=1}^m\widetilde m_a\,\hat\rho_a(x),\qquad
 \widetilde m_a=\sum_{j=1}^Kw_j\,n^{(j)}_a,\\
 \widetilde n_{2,w}(x,y)&=\sum_{a,b=1}^m\widetilde D_{ab}\,\hat\rho_a(x)\hat\rho_b(y),\quad
 \widetilde D_{ab}=\sum_{j=1}^Kw_jQ(\cfg^{(j)})_{ab},
\end{align}
\end{subequations}
where $\Sym$ averages over electron labels. The pair density $\widetilde n_{2,w}$ is therefore representable by construction. Its one-electron density is exactly $\rho$ whenever $\widetilde m=m$, and within a pair of cells it follows the product $\rho(x)\rho(y)$; spreading the electrons according to $\rho$, rather than uniformly, keeps $\rho$ exact at every level, so that only the pair density is approximated. For a representable $n_2$, the weights of the actual state reproduce its cell data, and the densities built from finer and finer partitions return to $n_2$ (Sec.~\ref{sec:constructive}). Representability on a partition thus becomes the question whether weights $w_j$ exist that reproduce the cell data $(m,D)$ of the given $n_2$.

\subsection{Occupation laws and the integer occupation polytope}
\label{sec:polytope}
The $K$ configurations of Eq.~\eqref{eq:nconfig} form the set
\begin{equation}
 \begin{split}
 \mathcal K_{N,m}&=\{\cfg^{(1)},\ldots,\cfg^{(K)}\}\\
 &=\Bigl\{\cfg=(n_1,\ldots,n_m):\ n_a\in\Z_{\ge0},\ \sum_{a=1}^mn_a=N\Bigr\}.
 \end{split}
 \label{eq:integersimplex}
\end{equation}
Because the cell counts are commuting operators with integer eigenvalues, a quantum state, pure or mixed, is with respect to these counts a probability-weighted combination of configurations; interference between different configurations does not enter. The weight $w_j$ is the probability of finding exactly $n^{(j)}_a$ electrons in every cell $a$, the quantity studied as electron number distribution function in real-space analyses of chemical bonding.\cite{Cances2004,Francisco2007,MartinPendas2007} The cell data of any $N$-electron state therefore satisfy
\begin{subequations}\label{eq:occupation_weights}
\begin{align}
 &\sum_{j=1}^Kw_j=1,\qquad
 \sum_{j=1}^Kw_j\,n^{(j)}_a=m_a,\\
 &\sum_{j=1}^Kw_j\,Q(\cfg^{(j)})_{ab}=D_{ab}
\end{align}
\end{subequations}
with $w_j\ge0$. Here $m$ and $D$ are fixed numbers and the $Q(\cfg)$ fixed integer matrices; only the weights are unknown. Geometrically, a definite configuration $\cfg$ has the cell data $(\cfg,Q(\cfg))$, a single point, and a state has a weighted average of these points, so the representable cell data form the polytope
\begin{equation}
 \begin{split}
 (m,D)\in\CN m&:=\conv\{(\cfg,Q(\cfg)):\cfg\in\mathcal K_{N,m}\}\\
 &\phantom{:}=\conv\bigl\{(k,kk^{\mathsf T}-\diag k):\\
 &\phantom{:=\conv\bigl\{}k\in\Z_{\ge0}^m,\ \1^{\mathsf T}k=N\bigr\},
 \end{split}
 \label{eq:integerpoly}
\end{equation}
where $\conv$ denotes the set of all weighted averages; its corners are exactly the configurations.\cite{FrieseckeVogler2018,Carlier2023} For a two-electron bond divided between two atomic cells $L$ and $R$, for example, $\cfg=(2,0)$ and $(0,2)$ are the ionic and $\cfg=(1,1)$ the covalent structure.\cite{MartinPendas2007} The ionic structures with equal weights give $m=(1,1)$, $D_{LL}=D_{RR}=1$, and $D_{LR}=0$; the covalent structure gives the same means with $D_{LL}=D_{RR}=0$ and $D_{LR}=1$; a stretched bond with both ionic and covalent weight lies between.

Testing a candidate pair density, for example that of a relaxed 2-RDM, is therefore a linear program (LP): find weights that satisfy Eq.~\eqref{eq:occupation_weights} for its cell data. If they exist, the candidate is consistent with integer electron numbers on this partition; if not, it cannot come from any $N$-electron state. With three electrons, for instance, a cell with mean $3/2$ requires $D_{LL}\ge1$, because the equal mixture of one and two electrons has the smallest $\langle\hat N_L(\hat N_L-1)\rangle$ among integer counts with this mean. The number $K$ of configurations, Eq.~\eqref{eq:nconfig}, grows rapidly with the number of cells, and all of them must be included: omitting configurations shrinks the polytope and can produce inequalities that physical states violate, whereas omitting cells only weakens the test (Table~\ref{tab:omission}). In the calculations reported here all configurations are enumerated, and every cut is checked on all of them; for larger problems, configurations can be generated on demand, provided that the final inequality is checked on all of them.

\begin{table}[tbp]
\caption{Effect of omissions on the computed bounds. Omitting configurations is the only one of the three that invalidates the result.}
\label{tab:omission}
\begin{ruledtabular}
\begin{tabular}{p{0.27\columnwidth}p{0.30\columnwidth}p{0.33\columnwidth}}
Omitted & Effect on the set & Effect on the result\\
\hline
$N$-representability conditions in a variational 2-RDM calculation & Feasible set enlarged (outer approximation) & Energy too low, but still a lower bound\\[3pt]
Configurations $\cfg$ in the occupation LP & Polytope $\CN m$ shrunk (inner approximation) & $\delta_{\mathcal P}$ too large; the cut may be violated by physical states\\[3pt]
Cells (coarser partition) & Fewer conditions on the pair density & $\delta_{\mathcal P}$ too small, but still a lower bound on $d_\rho$\\
\end{tabular}
\end{ruledtabular}
\end{table}

\subsection{The occupation LP: measuring a violation and finding a cut}
\label{sec:lp}
The reason for a failure is expressed by a quadratic function of the counts,
\begin{equation}
 q(\cfg)=c+\sum_{a=1}^mb_an_a+\sum_{a,b=1}^mW_{ab}Q(\cfg)_{ab},
 \label{eq:countquadratic}
\end{equation}
whose average over the cell data is obtained by summing over the configurations with the weights of Eq.~\eqref{eq:occupation_weights},
\begin{equation}
 \begin{split}
 \sum_{j=1}^Kw_j\,q(\cfg^{(j)})
 &=c\sum_{j=1}^Kw_j+\sum_{a=1}^mb_a\sum_{j=1}^Kw_jn^{(j)}_a\\
 &\quad+\sum_{a,b=1}^mW_{ab}\sum_{j=1}^Kw_jQ(\cfg^{(j)})_{ab}\\
 &=c+\sum_{a=1}^mb_am_a+\sum_{a,b=1}^mW_{ab}D_{ab}=:I[m,D].
 \end{split}
 \label{eq:finite_separator}
\end{equation}
If $q(\cfg)\ge0$ on every configuration, every $N$-electron state satisfies $I[m,D]\ge0$. To make the test quantitative, the equations for $D$ are allowed to fail, and the smallest failure is computed:
\begin{subequations}\label{eq:LPpair}
\begin{align}
 &\text{primal:}\ \ \delta_{\mathcal P}=\min_{w}\sum_{a,b=1}^m\Bigl|D_{ab}-\sum_{j=1}^Kw_jQ(\cfg^{(j)})_{ab}\Bigr|,\\
 &\qquad w_j\ge0,\ \ \sum_{j=1}^Kw_j=1,\ \ \sum_{j=1}^Kw_jn^{(j)}_a=m_a;\\
 &\text{dual:}\ \ \delta_{\mathcal P}=\max_{c,b,W}\bigl(-I[m,D]\bigr),\\
 &\qquad |W_{ab}|\le1,\ \ q(\cfg)\ge0\ \ (\cfg\in\mathcal K_{N,m}).
\end{align}
\end{subequations}
Both problems are linear, because the configurations and their pair counts are fixed numbers, and their optimal values are equal (standard form in the supplementary material). If $\delta_{\mathcal P}>0$, the dual solution is an inequality $I[m,D]\ge0$ that every $N$-electron state satisfies and the candidate violates by $\delta_{\mathcal P}$. It is linear in the 1- and 2-RDMs and can be imposed on the relaxation as a cut; Sec.~\ref{sec:constructive} shows that $\delta_{\mathcal P}$ is a lower bound on the distance from representability. These inequalities are finite versions of configuration-polynomial tests,\cite{Kuna2011} and $\langle(\hat N-k)(\hat N-k-1)\rangle\ge0$ is the simplest of them.

The two problems have a simple physical meaning:
\begin{itemize}
 \item The dual function $q$ is a classical lattice-gas Hamiltonian on the cells, with site energies $b_a$ and pair interactions $W_{ab}$, chosen by the LP and unrelated to the molecular Hamiltonian. Its ground-state energy over all configurations is nonnegative, so every $N$-electron state has $\langle q\rangle\ge0$. A candidate with $\langle q\rangle<0$ lies below the ground state of this cell Hamiltonian, the same kind of failure as a variational 2-RDM energy below the exact one, and the cut forbids it.
 \item Among all cell Hamiltonians with $|W_{ab}|\le1$, the dual selects the one that the candidate violates most strongly; the primal optimum is the smallest total change of the ordered-pair masses that makes the cell data consistent with integer electron numbers. Bounded interactions make the two equal.
 \item For two cells with a half-integer mean count and inferred variance $v<1/4$, each of the four entries of $D$ must move by $1/4-v$, so that $\delta=1-4v$ (Sec.~\ref{sec:gaussianrepair}).
 \item Finding the ground state of a general cell Hamiltonian is the classical Ising problem; this is the origin of the hardness discussed in Sec.~\ref{sec:molecules}.
\end{itemize}

\subsection{Two cells: the integer-count inequality}
\label{sec:twocell}
The simplest case is one region $\Omega$ and its complement. With
\begin{subequations}\label{eq:integercut}
\begin{align}
 \mu_\Omega&=\int_\Omega\rho,\qquad
 F_\Omega=\iint_{\Omega^2}n_2,\\
 I_{\Omega,k}&=F_\Omega-2k\mu_\Omega+k(k+1)\ge0
\end{align}
\end{subequations}
is the expectation of $(\hat N_\Omega-k)(\hat N_\Omega-k-1)$. If $\mu_\Omega=k+\theta$ with $0\le\theta<1$, it is the established integer-variance bound\cite{Yamada1961,Kuna2007,Kuna2011}
\begin{equation}
 v_\Omega:=F_\Omega+\mu_\Omega-\mu_\Omega^2
       \ge\theta(1-\theta):
 \label{eq:variancebound}
\end{equation}
a count with mean $5/2$ cannot have a variance below $1/4$. Equation~\eqref{eq:integercut} is the region cut imposed in molecular calculations in Part I [Eqs.~(2) and (3) there];\cite{NakataPartI} the quantities $\mu_\Omega$, $F_\Omega$, and $I_{\Omega,k}$ are the same in both papers. For two cells $\{\Omega,\Omega^c\}$ the LP reduces to Proposition~1 of Part I: $0\le\mu_\Omega\le N$, $I_{\Omega,k}\ge0$ for $k=0,\ldots,N-1$, and $F_\Omega\le(N-1)\mu_\Omega$. The $G$-type inequality of Garrod--Percus and Ayers--Davidson with $f=\chi_\Omega$ is the positivity $\langle(\hat N_\Omega-\mu_\Omega)^2\rangle\ge0$ and gives only $v_\Omega\ge0$.\cite{Garrod1964,AyersDavidson2006} It treats the count as a continuous quantity and does not use that its values are integers. Integrality raises the bound to $\theta(1-\theta)$: indeed $v_\Omega-\theta(1-\theta)=I_{\Omega,k}$, and the smallest variance at a given mean is reached by mixing the two neighboring integers $k$ and $k+1$. The difference matters most at a half-integer mean, where the bound is $1/4$ rather than $0$. This is the failure found in Part I for stretched N$_2^+$, whose fragments have the correct mean count $13/2$ but too small a fluctuation, and the gap between $0$ and $\theta(1-\theta)$ is exactly what the $G$ condition cannot see. For nonrepresentable data $v_\Omega$ is computed from $\mu_\Omega$ and $F_\Omega$ and is only an inferred variance, not the variance of an existing count; it can therefore lie below $\theta(1-\theta)$.

\subsection{Three or more cells: joint conditions}
\label{sec:threecells}
For three or more cells, the two-cell conditions are not sufficient, even when they hold for every region of space. Take $N=3$, three cells with $m=(1,1,1)$, $v=(1,1,-2)^{\mathsf T}$, and
\begin{equation}
 D=\begin{pmatrix}0&1&1\\1&0&1\\1&1&0\end{pmatrix}+\tfrac12vv^{\mathsf T}
  =\begin{pmatrix}1/2&3/2&0\\3/2&1/2&0\\0&0&2\end{pmatrix},
 \label{eq:threecell}
\end{equation}
with the cellwise pair density $\sum_{ab}D_{ab}\hat\rho_a(x)\hat\rho_b(y)$ of Eq.~\eqref{eq:coarsereconstruction} below. The first matrix is the pair table of the configuration $(1,1,1)$ with one electron in each cell (second row of Table~\ref{tab:threecell}): no pair within a cell and one pair between any two cells. The vector $v$ is chosen for three reasons. Its entries sum to zero, so the row sums of $D$, and with them the means $m$ and the contraction, are those of the configuration $(1,1,1)$. The matrix $vv^{\mathsf T}$ is positive semidefinite, so the perturbation can only increase the variance of the count of any region, and the two-cell tests, which detect too small a fluctuation, are passed. Its direction raises the pair number between cells 1 and 2 and lowers those between these cells and cell 3: with a coefficient $t$ in place of $\tfrac12$, $D_{12}=1+t$ and $D_{13}=D_{23}=1-2t$, so the three-cell inequality below is violated for $t>1/3$, nonnegativity requires $t\le1/2$, and $t=1/2$ gives the simple entries above. In detail, every region $\Omega$ passes the complete two-cell test: with $p_a=\int_\Omega \hat\rho_a$, the cellwise density of the configuration $(1,1,1)$, which is representable, has the same $\mu_\Omega$, and $F_\Omega$ exceeds its value by $(v^{\mathsf T}p)^2/2\ge0$, so every inequality $I_{\Omega,k}\ge0$ holds; the upper bound $F_\Omega\le2\mu_\Omega$ holds because the pair mass between $\Omega$ and its complement is nonnegative. The same positive semidefinite perturbation keeps the density covariance $\mathrm{Cov}$ of Eq.~\eqref{eq:kernel} positive semidefinite. Nevertheless no three-cell weights exist: every configuration satisfies $3n_1n_2\le2(n_1+n_2)$ (Table~\ref{tab:threecell}), so averaging with $m_1=m_2=1$ requires $D_{12}\le4/3$, whereas $D_{12}=3/2$. The distance is exact. Since the means fix the row sums, lowering $D_{12}$ and $D_{21}$ by at least $1/6$ each must be compensated elsewhere, so $\delta\ge2/3$; the equal mixture of the configurations $(0,0,3)$, $(1,2,0)$, and $(2,1,0)$ attains it, and since the candidate equals its own three-cell reconstruction, Eq.~\eqref{eq:coarseexact} gives $d_\rho(n_2)=\delta_3=2/3$, whereas $\delta=0$ for every partition into two regions. A violation invisible to every single-region test is thus detected by the joint three-cell LP with an exact finite distance (exact check in the supplementary material).

% Generated by verification/verify_three_cell.py; do not edit.
\begin{table}[tbp]
\caption{The ten configurations $(n_1,n_2,n_3)$ of three electrons in three cells and the inequality $3n_1n_2\le2(n_1+n_2)$ of the three-cell example. The coefficient $2/3$ is the largest ratio $n_1n_2/(n_1+n_2)$, attained at $(2,1,0)$ and $(1,2,0)$; averaging over any occupation law with $m_1=m_2=1$ gives $D_{12}\le4/3$.}
\label{tab:threecell}
\begin{ruledtabular}
\begin{tabular}{lccc}
Configurations & $n_1n_2$ & $n_1+n_2$ & $3n_1n_2$ vs.\ $2(n_1+n_2)$\\
\hline
(2,1,0), (1,2,0) & 2 & 3 & $6$ $=$ $6$\\
(1,1,1) & 1 & 2 & $3$ $<$ $4$\\
(3,0,0), (0,3,0) & 0 & 3 & $0$ $<$ $6$\\
(2,0,1), (0,2,1) & 0 & 2 & $0$ $<$ $4$\\
(1,0,2), (0,1,2) & 0 & 1 & $0$ $<$ $2$\\
(0,0,3) & 0 & 0 & $0$ $=$ $0$\\
\end{tabular}
\end{ruledtabular}
\end{table}

\subsection{Complete diagnosis through refinement}
A single partition gives only necessary conditions. Refinement removes this limitation: counting the electrons in ever finer cells characterizes the representable pair densities completely. The intuitive reason is that the probability distribution of the electron positions is approximated arbitrarily well by the distribution of the electron counts in sufficiently fine cells.
\begin{theorem}[Partition characterization; known result\cite{KonstantopoulosYuan2019,Carlier2023}]
\label{thm:partition}
A pair density $n_2$ with one-electron density $\rho$ is the pair density of a probability distribution of $N$ electrons, symmetric in the electron labels, if and only if for every finite partition of space into cells $A_a$ the cell data $m_a=\int_{A_a}\rho$ and $D_{ab}=\int_{A_a}\int_{A_b}n_2$ satisfy Eq.~\eqref{eq:integerpoly}. One sequence of successively finer partitions suffices, provided that its cells become arbitrarily small. Cell boundaries such as planes and spheres have zero volume and therefore carry no electrons.
\end{theorem}
\begin{proof}[Proof sketch]
Necessity is the counting argument of Sec.~\ref{sec:polytope}. Sufficiency is constructive. On each partition, electrons are placed into the cells with the configuration probabilities of the LP and distributed inside each cell according to $\rho$; this gives an actual $N$-electron distribution with the same cell data and the one-electron density $\rho$. Because $\rho$ is the same at every level, no electron escapes to infinity along the sequence, so a subsequence of these distributions converges as the cells shrink. Since the cell boundaries carry no electrons, the limit keeps the prescribed pair mass of every cell pair, and since the cells become arbitrarily small, its pair density is $n_2$; the limit need not be unique. The complete proof is given in the supplementary material.
\end{proof}
The theorem is a consequence of exchangeability and extension theory; it is stated here in the present notation. With spin, space is replaced by $\R^3\times\{\uparrow,\downarrow\}$ and the statement is unchanged (Sec.~\ref{sec:spin}). For a fixed number of particles, Kuna, Lebowitz, and Speer state the equivalent dual form: a pair $(\rho,n_2)$ is realizable if and only if $f_0+\int f_1\rho+\iint f_2n_2\ge0$ for every quadratic configuration function $f_0+\sum_if_1(x_i)+\sum_{i\ne j}f_2(x_i,x_j)$ that is nonnegative on all $N$-point configurations (Corollary~3.10 of Ref.~\citenum{Kuna2011}, which they attribute to Percus); the dual problem of Eq.~\eqref{eq:LPpair} is this condition with $f_1$ and $f_2$ constant on the cells. Here $f_1$ is a one-body potential and $f_2$ a pair potential, and the configuration function is the energy of $N$ classical particles in them. With $f_1(x)=b_a$ for $x$ in cell $a$ and $f_2(x,y)=W_{ab}$ for $x$ in cell $a$ and $y$ in cell $b$, the energy depends on the positions only through the cell counts and equals $q(\cfg)$ of Eq.~\eqref{eq:countquadratic}, so that the condition on all point configurations reduces to the $K$ configurations, and the two integrals become $\sum_ab_am_a$ and $\sum_{ab}W_{ab}D_{ab}$, which gives $I[m,D]$ of Eq.~\eqref{eq:finite_separator}. As the cells shrink, step functions approximate any potentials: the site energies $b_a$ and cell interactions $W_{ab}$ become $f_1$ and $f_2$, while the cell data $m$ and $D$ become $\rho$ and $n_2$, as a potential and a density are paired in density-functional theory. The condition is the classical, diagonal counterpart of the Garrod--Percus duality between $N$-representable 2-RDMs and two-body Hamiltonians\cite{Garrod1964} (Table~\ref{tab:duality}).

\begin{table*}[tbp]
\caption{Representability and Hamiltonian duality: the quantum condition of Garrod and Percus, its classical counterpart for the pair density, and the finite cell version solved by the occupation LP.}
\label{tab:duality}
\begin{ruledtabular}
\begin{tabular}{p{0.15\textwidth}p{0.26\textwidth}p{0.26\textwidth}p{0.26\textwidth}}
 & Garrod--Percus (quantum)\cite{Garrod1964} & Kuna--Lebowitz--Speer (classical)\cite{Kuna2011} & Occupation LP (cells)\\
\hline
Object & 2-RDM $\Gamma$, including off-diagonal elements & Pair density $n_2$ with $\rho$ & Cell data $(m,D)$\\[3pt]
Test Hamiltonian & Any two-body operator $H$; terms need not commute & Classical energy $F$ with one-body potential $f_1$ and pair potential $f_2$ & Cell Hamiltonian $q(\cfg)$ with site energies $b_a$ and cell interactions $W_{ab}$\\[3pt]
Representable if and only if & $\Tr(\Gamma H)\ge E_0(H)$ for every $H$ & $f_0+\int f_1\rho+\iint f_2n_2\ge0$ for every $F\ge0$ on all point configurations & $I[m,D]\ge0$ for every $q\ge0$ on the $K$ configurations\\[3pt]
Ground-state problem & Quantum eigenvalue problem & Minimum over positions of $N$ particles & Minimum over $K$ configurations\\[3pt]
Practical subfamilies & Sums of squares: $P$, $Q$, $G$, $T_1$, $T_2$, $(2,q)$-positivity & Count inequalities on regions & Complete on a fixed partition; dual solutions are the cuts\\
\end{tabular}
\end{ruledtabular}
\end{table*}
In particular, a pair density that is not representable is detected by the LP of some finite partition, whose dual solution certifies the violation. A pass on finitely many partitions, however, does not prove representability. This motivates the quantitative result below, which measures how far a pair density lies from the representable set rather than only testing it.

The partition conditions form a hierarchy of relaxations that tightens monotonically. For a finite partition $\mathcal P$ with $|\mathcal P|$ cells, let $\mathcal R_{\mathcal P}$ denote the set of pair densities with contraction $\rho$ whose cell data satisfy Eq.~\eqref{eq:integerpoly}, $(m,D)\in\CN{|\mathcal P|}$.
\begin{proposition}[Monotone hierarchy of relaxations; (i)--(iii) from realizability and extension theory]
\label{prop:hierarchy}
Let $\mathcal P_1,\mathcal P_2,\dots$ be finite partitions in which $\mathcal P_{\ell+1}$ refines $\mathcal P_\ell$, and write $\mathcal R_\ell=\mathcal R_{\mathcal P_\ell}$, with all integer configurations of each level included.
\begin{enumerate}
 \item[(i)] Every representable pair density with contraction $\rho$ lies in $\mathcal R_\ell$ for every $\ell$.
 \item[(ii)] $\mathcal R_{\ell+1}\subseteq\mathcal R_\ell$.
 \item[(iii)] If the cells of the sequence become arbitrarily small (they generate the Borel sets), $\bigcap_\ell\mathcal R_\ell$ is the set of pair densities with contraction $\rho$ that have a symmetric probability-measure extension (Theorem~\ref{thm:partition}), the $L^1$ closure of $\mathcal P^{\mathrm{ac}}_{N,\rho}$ (supplementary material), and the LP optima increase to the distance, $\delta_\ell\uparrow d_\rho(n_2)$ (Theorem~\ref{thm:distance}).
\end{enumerate}
In a finite orbital basis, let $\mathcal C$ be a closed and bounded set of pairs $(\gamma,\Gamma)$ that contains those of all $N$-electron ensembles, for example the $P$, $Q$, $G$ (and $T_1$, $T_2$) conditions with traces and contractions, and let $E_\ell$ be the minimum of the energy over $(\gamma,\Gamma)\in\mathcal C$ whose cell data, Eq.~\eqref{eq:pullback}, satisfy the level-$\ell$ conditions. Then
\begin{equation}
 E_\ell\le E_{\ell+1}\le E_0,\qquad E_\ell\uparrow E_\infty,
 \label{eq:energyhierarchy}
\end{equation}
where $E_0$ is the full configuration-interaction energy in the same finite basis and $E_\infty$ is the minimum over $(\gamma,\Gamma)\in\mathcal C$ whose cell data satisfy the conditions of every level.
\end{proposition}
\begin{proof}[Proof sketch]
(i) Counting the electrons of an actual state in the cells gives an occupation law. (ii) Merging cells merges configurations: an occupation law on a finer partition, read on the coarser one, is an occupation law there, so a finer partition imposes every condition of a coarser one. (iii) combines Theorem~\ref{thm:partition} with Theorem~\ref{thm:distance} below. For the energies, every $N$-electron ensemble, in particular the exact ground state in the basis, satisfies the conditions of every level, so each $E_\ell$ is a lower bound to $E_0$; adding conditions can only raise the minimum; and because $\mathcal C$ is bounded, the minima converge to the minimum over the conditions of all levels. The complete proof is given in the supplementary material.
\end{proof}
Parts (i)--(iii) are known;\cite{Kuna2007,Kuna2011,KonstantopoulosYuan2019,Carlier2023} the proposition adds the energy statement, Eq.~\eqref{eq:energyhierarchy}. Because consistency with the 1-RDM is not imposed, in general $E_\infty<E_0$;\cite{Percus2005,LiuPercus2006} this separate consistency problem is discussed in Sec.~\ref{sec:onerdm}. Imposing only some conditions of a level, such as finitely many dual witnesses in Sec.~\ref{sec:molecules}, keeps the lower bound; omitting configurations does not (Table~\ref{tab:omission}). Strict-gap examples and a comparison with prior work are given in the supplementary material.

\section{Constructive approximation and a convergent distance bracket}
\label{sec:constructive}
\subsection{Coarse reconstruction and coarse-graining error}
A partition sees only the cell-pair masses $D_{ab}$ of a pair density. The simplest pair density with these masses spreads each of them inside its pair of cells according to $\rho(x)\rho(y)$, as a histogram replaces a curve by its bin averages. This coarse reconstruction (the lift of Eq.~\eqref{eq:densitylift} with the candidate's own $D$)\cite{CarlierDuval2017} and its error are
\begin{equation}
 n_{2,\mathcal P}(x,y)=\sum_{a,b=1}^mD_{ab}\hat\rho_a(x)\hat\rho_b(y),\quad
 e_{\mathcal P}=\|n_2-n_{2,\mathcal P}\|_1.
 \label{eq:coarsereconstruction}
\end{equation}
It keeps $\rho$ and all cell-pair masses, but it need not be representable, because $D$ need not be an average of configuration tables. The coarse-graining error $e_{\mathcal P}$ is what the partition cannot see. The lift $\widetilde n_{2,w}$ of an occupation law, by contrast, is representable by construction, and because the cells do not overlap and each $\hat\rho_a$ integrates to one, the distance between the two lifts is the distance between their tables: on each bin $A_a\times A_b$ the two lifts differ by $(\widetilde D_{ab}-D_{ab})\hat\rho_a(x)\hat\rho_b(y)$, so that
\begin{equation}
 \begin{split}
 &\|\widetilde n_{2,w}-n_{2,\mathcal P}\|_1\\
 &=\sum_{a,b=1}^m\iint_{A_a\times A_b}\bigl|\widetilde n_{2,w}(x,y)-n_{2,\mathcal P}(x,y)\bigr|\dd x\dd y\\
 &=\sum_{a,b=1}^m|\widetilde D_{ab}-D_{ab}|\iint_{A_a\times A_b}\hat\rho_a(x)\hat\rho_b(y)\dd x\dd y\\
 &=\sum_{a,b=1}^m|\widetilde D_{ab}-D_{ab}|.
 \end{split}
 \label{eq:liftisometry}
\end{equation}
A repair of the finite table is therefore a repair of the pair density.

\subsection{The fixed-$\rho$ occupation LP}
Define
\begin{equation}
 \delta_{\mathcal P}
 =\min_{\substack{w_j\ge0,\ \sum_{j=1}^Kw_j=1\\ \sum_{j=1}^Kw_j\cfg^{(j)}=m}}
       \left\|D-\sum_{j=1}^Kw_jQ(\cfg^{(j)})\right\|_{1,\mathrm{entry}},
 \label{eq:deltaLP}
\end{equation}
the smallest change of the cell-pair masses, summed over all ordered entries, that makes them consistent with integer electron numbers at the given means. It is the primal problem of Eq.~\eqref{eq:LPpair}, a finite LP, and $\delta_{\mathcal P}=0$ exactly when the cell data are representable.

The next theorem turns the yes-or-no test of Theorem~\ref{thm:partition} into a measurement. Here $d_\rho(n_2)$ of Eq.~\eqref{eq:distance_definition} is the smallest change of the pair density, with $\rho$ kept fixed, that makes it representable. In words: a violation found on a partition is real and never larger than the true distance; the true distance exceeds it at most by $e_{\mathcal P}$; and as the cells shrink, $e_{\mathcal P}$ vanishes, so that the LP optimum rises to the true distance.
\begin{theorem}[Constructive distance bounds and convergence]
\label{thm:distance}
Let $n_2$ satisfy Sec.~\ref{sec:definitions}, with its contraction $\rho$ fixed. For every finite partition,
\begin{equation}
 \boxed{\delta_{\mathcal P}\le d_\rho(n_2)
             \le\delta_{\mathcal P}+e_{\mathcal P},}
 \label{eq:distancebracket}
\end{equation}
and the lift of an optimal occupation law, Eq.~\eqref{eq:densitylift}, is a representable pair density within the upper bound. Along successively finer partitions whose cells become arbitrarily small,
\begin{equation}
 \delta_\ell\le\delta_{\ell+1},\qquad e_\ell\longrightarrow0,
 \qquad \delta_\ell\uparrow d_\rho(n_2).
 \label{eq:conditionalconvergence}
\end{equation}
For the coarse reconstruction itself the LP is exact:
\begin{equation}
 \boxed{d_\rho(n_{2,\mathcal P})=\delta_{\mathcal P}.}
 \label{eq:coarseexact}
\end{equation}
\end{theorem}
\begin{proof}[Proof sketch]
Lower bound: the cell data of any representable pair density are feasible in Eq.~\eqref{eq:deltaLP}, and integrating over cells can only reduce a difference. Upper bound: the lift of an optimal law is representable and lies at distance $\delta_{\mathcal P}$ from $n_{2,\mathcal P}$, Eq.~\eqref{eq:liftisometry}, which lies at distance $e_{\mathcal P}$ from $n_2$; the same facts applied to $n_{2,\mathcal P}$ give Eq.~\eqref{eq:coarseexact}. Monotonicity: merging cells maps every fine repair to a coarse one that is no more expensive. Convergence: $e_{\mathcal P}\to0$, Eq.~\eqref{eq:ehbin} below. The complete proof is given in the supplementary material.
\end{proof}

\emph{What the coarse-graining error measures.} The reconstruction keeps the shape $\rho(x)\rho(y)$ within each bin $A_a\times A_b$, so what it averages is the ratio $n_2/(\rho\rho)$, the pair correlation function $h(x,y)$, which equals one for uncorrelated electrons:
\begin{subequations}\label{eq:conditionalratio}
\begin{align}
 h(x,y)&:=\frac{N^2}{M}\,\frac{n_2(x,y)}{\rho(x)\rho(y)}\quad(\rho(x)\rho(y)>0),\\
 \bar h_{ab}&=\frac{\iint_{A_a\times A_b}h(x,y)\rho(x)\rho(y)\dd x\dd y}{\iint_{A_a\times A_b}\rho(x)\rho(y)\dd x\dd y}.
\end{align}
\end{subequations}
Inserting $\hat\rho_a=\rho\chi_a/m_a$ into Eq.~\eqref{eq:coarsereconstruction}, with $D_{ab}=(M/N^2)\iint_{A_a\times A_b}h\rho\rho$ and $\iint_{A_a\times A_b}\rho\rho=m_am_b$, gives $n_{2,\mathcal P}(x,y)=(M/N^2)\,\bar h_{ab}\,\rho(x)\rho(y)$ on the bin. On each bin $A_a\times A_b$ the candidate and its reconstruction therefore differ by $(M/N^2)\,[h(x,y)-\bar h_{ab}]\,\rho(x)\rho(y)$, so that
\begin{equation}
 \begin{split}
 e_{\mathcal P}&=\sum_{a,b=1}^m\iint_{A_a\times A_b}\bigl|n_2(x,y)-n_{2,\mathcal P}(x,y)\bigr|\dd x\dd y\\
 &=\sum_{a,b=1}^m\iint_{A_a\times A_b}\Bigl|\frac{M}{N^2}\bigl[h(x,y)-\bar h_{ab}\bigr]\rho(x)\rho(y)\Bigr|\dd x\dd y\\
 &=\frac{M}{N^2}\sum_{a,b=1}^m\iint_{A_a\times A_b}|h(x,y)-\bar h_{ab}|\,\rho(x)\rho(y)\dd x\dd y.
 \end{split}
 \label{eq:ehbin}
\end{equation}
For uncorrelated electrons, $h\equiv1$, the reconstruction is exact on every partition; only the variation of $h$ within a bin contributes. The bin averages approach $h$ as the cells shrink, as a histogram approaches a curve when the bins shrink, so $e_{\mathcal P}\to0$ for every integrable $h$, smooth or not; smoothness enters only the rate, Eq.~\eqref{eq:ratebound}.

\subsection{Dual witnesses and rigorous finite computation}
The dual problem of Eq.~\eqref{eq:LPpair} reads
\begin{subequations}\label{eq:LPdual}
\begin{align}
 \delta_{\mathcal P}={}&\max_{c,b,W}\Bigl\{-c-\sum_{a=1}^mb_am_a-\sum_{a,b=1}^mW_{ab}D_{ab}\Bigr\},\\
 &|W_{ab}|\le1,\\
 &q(\cfg^{(j)})\ge0\quad(j=1,\ldots,K),
\end{align}
\end{subequations}
with $q$ the cell Hamiltonian of Eq.~\eqref{eq:countquadratic}. It is the condition of Kuna, Lebowitz, and Speer restricted to step-function potentials, $f_1=b_a$ and $f_2=W_{ab}$ on the cells, with the normalization $|W_{ab}|\le1$ that turns their sign test into a measurement: its optimum equals the primal optimum $\delta_{\mathcal P}$. A positive optimum gives the cell Hamiltonian $q(\cfg)$ of Eq.~\eqref{eq:countquadratic}, nonnegative on every configuration and negative on the candidate; it is the cut added to the SDP in Sec.~\ref{sec:molecules}. Any feasible point of the dual LP gives a lower bound to the distance, and any occupation law an upper bound, so the LP need not be solved to optimality. The model examples, including an exact rational benchmark in the supplementary material, are checked in exact arithmetic; the molecular values of Sec.~\ref{sec:molecules} are numerical (conditions for certified values in the supplementary material).

\subsection{Two directions of refinement and the cost of resolution}
The relaxations can be tightened in two directions: by stronger conditions on a fixed partition, from single-region counts to the complete occupation polytope, and by finer partitions. Theorem~\ref{thm:distance} concerns the second direction with the complete LP at each level. When both are truncated, every valid condition found so far should be kept, and no polynomial cost is claimed. On all of $\R^3$ an exterior cell must remain part of the partition, so that no electron is dropped.

How fine the cells must be is set by the pair correlation function $h(x,y)$ of Eq.~\eqref{eq:conditionalratio}: the coarse-graining error is small once $h$ changes little across a pair of cells. Let the cells have diameter at most $\eta$ inside a region $C$ around the molecule, with one exterior cell for the rest of space; let $\omega(t)$ bound the change of $h$ when its two arguments move within $C$ by a total distance $t$ [$\omega(t)=Lt$ if $h$ has slope at most $L$]; and let $\tau=\int_{X\setminus C}\rho$ be the number of electrons outside $C$. Then
\begin{equation}
 e_{\mathcal P}\le
 \underbrace{M\omega(2\eta)}_{\text{averaging inside }C}
 +\underbrace{4(N-1)\tau}_{\text{exterior cell}}.
 \label{eq:ratebound}
\end{equation}
Both terms follow from Eq.~\eqref{eq:ehbin}. For a bin of two cells inside $C$, two of its points differ by at most $\eta$ in each argument, so $|h-\bar h_{ab}|\le\omega(2\eta)$ there, and since $\iint_{C\times C}\rho\rho\le N^2$ these bins contribute at most $M\omega(2\eta)$. For a bin that contains the exterior cell, $|n_2-n_{2,\mathcal P}|\le n_2+n_{2,\mathcal P}$; by the contraction, the pair mass of $n_2$ with $x$ outside $C$ is $(N-1)\tau$, and the same holds with $y$ outside $C$, so these bins carry at most $2(N-1)\tau$ of $n_2$, and the same of $n_{2,\mathcal P}$, which has the same bin masses. Enlarging $C$ reduces $\tau$ but takes in the tails of the density, where $h$, a ratio of small densities, may vary rapidly; shrinking $\eta$ reduces the first term but enlarges the LP. The bound requires an estimate of the variation of $h$, which is not available for a relaxed 2-RDM, so it does not by itself give an algorithm for choosing partitions.

\subsection{Spin-resolved pair densities at fixed \texorpdfstring{$S_z$}{Sz}}
\label{sec:spin}
Spin labels matter when a molecule dissociates into open-shell fragments: a stretched Li$_2$ molecule must give each doublet atom $S_z=\pm\tfrac12$, whereas a relaxation can assign each atom a fractional spin. This subsection shows which such errors the spin-resolved counts detect. Electronic pair densities carry spin labels, and the occupation theory uses them directly. Let $N=N_\uparrow+N_\downarrow$, and let a spin-resolved candidate consist of nonnegative blocks $n_2^{\sigma\tau}(x,y)=n_2^{\tau\sigma}(y,x)$, $\sigma,\tau\in\{\uparrow,\downarrow\}$, with
\begin{subequations}\label{eq:spinblocks}
\begin{align}
 \iint n_2^{\sigma\tau}&=N_\sigma(N_\tau-\delta_{\sigma\tau}),\\
 \int n_2^{\sigma\tau}(x,y)\dd y&=(N_\tau-\delta_{\sigma\tau})\rho_\sigma(x).
\end{align}
\end{subequations}
Its spin sum $n_2=\sum_{\sigma\tau}n_2^{\sigma\tau}$ is an ordinary pair density with $\rho=\rho_\uparrow+\rho_\downarrow$. Representability now asks for one $N$-electron distribution with exactly $N_\uparrow$ up-spin electrons that reproduces all four blocks. On a partition, a configuration consists of the up-spin and down-spin counts $k^\sigma=(k^\sigma_1,\ldots,k^\sigma_m)$ of all cells, with $\sum_{a=1}^mk^\sigma_a=N_\sigma$ for $\sigma=\uparrow,\downarrow$, and its pair tables are
\begin{equation}
 Q^{\sigma\tau}(k)_{ab}=k^\sigma_ak^\tau_b-\delta_{\sigma\tau}\delta_{ab}k^\sigma_a\qquad(a,b=1,\ldots,m),
 \label{eq:spinQ}
\end{equation}
the numbers of ordered pairs with a $\sigma$ electron in cell $a$ and a $\tau$ electron in cell $b$. The spin-resolved LP optimum $\delta^{\mathrm{spin}}_{\mathcal P}$ and distance $d^{\mathrm{spin}}$ are defined as before, with both spin densities fixed and the four blocks summed.

\begin{proposition}[Spin resolution at fixed $S_z$]
\label{prop:spin}
(i) Theorems~\ref{thm:partition} and~\ref{thm:distance} hold for spin-resolved candidates, with the configurations $(k^\uparrow,k^\downarrow)$ and the cell densities $\hat\rho^\sigma_a=\rho_\sigma\chi_a/m^\sigma_a$.
(ii) Spin summation maps spin-resolved representable candidates to representable pair densities, and $d_\rho(n_2)\le d^{\mathrm{spin}}$.
(iii) The converse fails: a representable spin-summed pair density can have spin blocks at positive spin-resolved distance.
\end{proposition}
\begin{proof}[Proof sketch]
(i) The previous arguments apply with the spin label treated as part of the cell: cells $A_a\times\{\sigma\}$, lifts built species by species, and the number of up-spin electrons kept fixed in every step and in the limit (supplementary material). (ii) Forgetting the spin labels maps an extension to an extension and can only shorten distances. (iii) See the examples below.
\end{proof}

The first example shows what spin summation hides. It mimics a stretched Li$_2$ molecule, whose two doublet atoms each hold three electrons: a physical state gives each atom $S_z=\pm\tfrac12$, so that the up-spin count of one atom is 1 or 2, whereas the candidate below assigns each atom $3/2$ up-spin electrons without fluctuation. Take two cells $\{\Omega,\Omega^c\}$, $N_\uparrow=N_\downarrow=3$, $\rho_\uparrow=\rho_\downarrow$ with $\int_\Omega\rho_\sigma=3/2$, and the cellwise blocks with tables
\begin{equation}
 D^{\sigma\sigma}=\begin{pmatrix}3/4&9/4\\9/4&3/4\end{pmatrix},\qquad
 D^{\uparrow\downarrow}=D^{\downarrow\uparrow}=\frac94\begin{pmatrix}1&1\\1&1\end{pmatrix}.
 \label{eq:spinexample}
\end{equation}
These are obtained by inserting, in place of an integer configuration, the fractional configuration $k^\uparrow=k^\downarrow=(3/2,3/2)$, three halves of an electron of each spin in each cell without fluctuation, into the pair tables of Eq.~\eqref{eq:spinQ}: $\tfrac94-\tfrac32=\tfrac34$ and $\tfrac94$ for the same spin, and $\tfrac94$ for opposite spins.
Let $U$ be the number of up-spin electrons in $\Omega$. The same-spin table gives its mean, $\langle U\rangle=3/2$, and $\langle U(U-1)\rangle=D^{\uparrow\uparrow}_{\Omega\Omega}=3/4$; its rows sum to $(N_\sigma-1)m^\sigma_a=3$, as the contraction requires. The opposite-spin table is the product $m^\uparrow_am^\downarrow_b=9/4$ of the means, as for up- and down-spin counts that are uncorrelated. Summing the four blocks gives the spin-summed table $D=D^{\uparrow\uparrow}+D^{\downarrow\downarrow}+D^{\uparrow\downarrow}+D^{\downarrow\uparrow}=\bigl(\begin{smallmatrix}6&9\\9&6\end{smallmatrix}\bigr)$, which is the table $Q(3,3)$ of exactly three electrons in each cell, so the spin-summed pair density is representable. The up-spin count, however, violates the integer-variance bound. Every count with mean $\mu_\Omega=k+\theta$ satisfies Eq.~\eqref{eq:variancebound},
\begin{subequations}\label{eq:spinvar}
\begin{equation}
 F_\Omega+\mu_\Omega-\mu_\Omega^2\ge\theta(1-\theta),
\end{equation}
whereas for $U$, with $F_\Omega=D^{\uparrow\uparrow}_{\Omega\Omega}$, $\mu_\Omega=m^\uparrow_\Omega=3/2$, and $\theta=1/2$,
\begin{equation}
 D^{\uparrow\uparrow}_{\Omega\Omega}+m^\uparrow_\Omega-\bigl(m^\uparrow_\Omega\bigr)^2
 =\tfrac34+\tfrac32-\bigl(\tfrac32\bigr)^2=0<\tfrac14 .
\end{equation}
\end{subequations}
At mean $3/2$ the bound requires $D^{\uparrow\uparrow}_{\Omega\Omega}\ge1$. Raising the same-spin entry from $3/4$ to $1$ moves each of the four entries of a same-spin table by $1/4$, a cost of 1 per spin, so that
\begin{equation}
 d_\rho(n_2)=0,\qquad d^{\mathrm{spin}}=\delta^{\mathrm{spin}}_{\{\Omega,\Omega^c\}}=2.
 \label{eq:spindistance}
\end{equation}
Each cell holds exactly three electrons but $3/2$ up-spin electrons without fluctuation, a fractional-spin population, the spin counterpart of a fractional charge.\cite{Cohen2008} It passes every test on the spin-summed pair density and the spin-resolved $G$-type conditions of Ayers and Davidson;\cite{AyersDavidson2006} the single-species count of either spin, imposed in Part I, detects it.

Some candidates pass the charge, up-spin, and down-spin tests separately and are nevertheless not representable; only the joint law of both spins detects them. The second example is of this kind. It mimics a stretched LiH-like molecule with four electrons, the cell $L$ containing the hydrogen end with one electron on average; $L$ then holds H$^+$, H$\uparrow$, H$\downarrow$, or H$^-$, and a physical state assigns each of these a nonnegative probability. Take two cells $L$, $R$, $N_\uparrow=N_\downarrow=2$, let $U$ and $V$ be the up- and down-spin counts in $L$, and prescribe $\langle U\rangle=1/4$, $\langle V\rangle=3/4$, $\langle U(U-1)\rangle=\langle V(V-1)\rangle=0$, and $\langle UV\rangle=5/16$. Since $\langle U(U-1)\rangle=0$ and $U$ is a nonnegative integer, $U$ takes only the values 0 and 1, with $\Pr(U=1)=\langle U\rangle=1/4$: the up-spin electron of $L$ is present with probability $1/4$. Likewise $V\in\{0,1\}$ with $\Pr(V=1)=3/4$. Each species alone is thus a valid law. So is the spin sum $S=U+V$, with mean 1 and $\langle S(S-1)\rangle=2\langle UV\rangle=5/8$, which the law $\Pr(S=0,1,2)=(5/16,3/8,5/16)$ reproduces; every spatial partition therefore passes all three separate tests. The joint law, however, is impossible. The prescribed moments fix the four probabilities completely,
\begin{equation}
 \begin{split}
 \Pr(\mathrm H^-)&=\langle UV\rangle=\tfrac5{16},\qquad
 \Pr(\mathrm H{\uparrow})=\langle U\rangle-\langle UV\rangle=-\tfrac1{16},\\
 \Pr(\mathrm H{\downarrow})&=\langle V\rangle-\langle UV\rangle=\tfrac7{16},\qquad
 \Pr(\mathrm H^+)=\tfrac5{16},
 \end{split}
 \label{eq:hprob}
\end{equation}
and one of them is negative: the weight of H$^-$, with both spins on hydrogen, exceeds the probability $1/4$ of an up-spin electron there. As a count inequality, the integer $U-V$ gives
\begin{equation}
 \tfrac12\langle(U-V)(U-V+1)\rangle\ge0,
 \label{eq:jointspincut}
\end{equation}
whose candidate value is $\langle U\rangle-\langle UV\rangle=-1/16$; it is the count inequality of the spin-labeled region $(L,\uparrow)\cup(R,\downarrow)$, and $d^{\mathrm{spin}}=1/2$. This candidate also passes the spin-resolved $G$-type conditions; only integrality excludes it. The derivations and exact checks of both examples are given in the supplementary material.

In Part I,\cite{NakataPartI} the candidate cuts of Eq.~\eqref{eq:integercut} for one region included the spin-summed, up-spin, and down-spin counts; the last two are single-species two-cell members of this family. The counts constrain only how many up-spin and down-spin electrons are in each region. The local spin of a fragment, $\hat{\bm S}_A^2=\hat S_{Az}^2+\tfrac12(\hat S_{A+}\hat S_{A-}+\hat S_{A-}\hat S_{A+})$, also contains spin-flip terms, which no condition on the counts controls (Sec.~\ref{sec:remaining}). Because Proposition~\ref{prop:spin} is formulated at fixed $S_z$, it does not address the dependence of approximate relaxations on the $M_S$ component of a multiplet\cite{Nakata2008,VanAggelen2012} or on the projection of the orbital angular momentum.\cite{LiDePrince2019} The explicit constructions of Secs.~\ref{sec:gaussianinputs} and~\ref{sec:gaussian} use one common spin, so for them the spin-resolved and spin-summed problems coincide.

\section{Pair densities from orbital-basis 2-RDMs}
\label{sec:gaussianinputs}
In a variational 2-RDM calculation the pair density is not a free function: it is built from a 1-RDM and a 2-RDM in a finite orbital basis. This section states how the real-space cell data and cuts are evaluated from these matrices and what the $G$ condition already guarantees.
\subsection{Finite-orbital moment data and the pair-density map}
A relaxed 2-RDM need not come from an $N$-electron state, so we treat its elements as candidate moments: values of a normalized linear functional $\Lcal$ on products of creation and annihilation operators, with $\Lcal(X)=\langle X\rangle$ for a genuine state. Let $2\le N\le r$ and
\begin{subequations}\label{eq:rdms}
\begin{align}
 \gamma_{pr}&=\Lcal(a_p^\dagger a_r),\\
 \Gamma_{pq;rs}&=\tfrac12\Lcal(a_p^\dagger a_q^\dagger a_s a_r),
\end{align}
\end{subequations}
with the usual antisymmetries and Hermiticity, and
\begin{equation}
 \Lcal(1)=1,\quad \Tr\gamma=N,\quad
 \sum_q\Gamma_{pq;rq}=\tfrac12(N-1)\gamma_{pr}.
 \label{eq:contraction}
\end{equation}
This is the convention of Part I,\cite{NakataPartI} with trace $N(N-1)/2$; the pair density $n_2$ and the regional quantities of Eq.~\eqref{eq:integercut} are the same in both papers. For orthonormal orbitals $\phi_i(x)$,
\begin{subequations}\label{eq:pairmap}
\begin{align}
 \rho(x)&=\sum_{pr}\gamma_{pr}\phi_p^*(x)\phi_r(x),\\
 n_2(x,y)&=2\sum_{pqrs}\Gamma_{pq;rs}
 \phi_p^*(x)\phi_q^*(y)\phi_r(x)\phi_s(y),
\end{align}
\end{subequations}
so that $\int n_2(x,y)\dd y=(N-1)\rho(x)$. All explicit constructions below have a single common spin.

The cell data need only the overlap matrices of the orbitals over the cells,
\begin{subequations}\label{eq:brackets}
\begin{align}
 S^a_{pr}&=\int_{A_a}\phi_p^*(x)\phi_r(x)\dd x,\\
 \gfun R&=\sum_{pr}\gamma_{pr}R_{pr},\\
 \Gamma[R,T]&=2\sum_{pqrs}\Gamma_{pq;rs}R_{pr}T_{qs},
\end{align}
\end{subequations}
which give
\begin{subequations}\label{eq:pullback}
\begin{align}
 m_a&=\gfun{S^a},\qquad D_{ab}=\Gamma[S^a,S^b],\\
 0&\preceq S^a\preceq I,\qquad \sum_aS^a=I.
\end{align}
\end{subequations}
The $S^a$ are neither projections nor mutually commuting. The operator $\sum_{pr}S^a_{pr}a_p^\dagger a_r$ that counts electrons of cell $a$ within the orbital space therefore does not have integer eigenvalues, and positivity conditions on the 2-RDM do not see the integrality of the real-space count. A cut of the form of Eq.~\eqref{eq:finite_separator} becomes
\begin{equation}
 c+\sum_a b_a\gfun{S^a}+\sum_{ab}W_{ab}\Gamma[S^a,S^b]\ge0,
 \label{eq:fixedcut}
\end{equation}
which is linear in the 2-RDM after contraction eliminates $\gamma$; it is added to the SDP in Sec.~\ref{sec:molecules} without a change of orbital basis. Partial-overlap contractions of this kind are established tools.\cite{Chuiko2026}

\subsection{What the \texorpdfstring{$G$}{G} condition already guarantees}
With $B=\sum_{p<q}b_{pq}a_qa_p$ and $A=\sum_{pq}A_{pq}a_p^\dagger a_q$, the usual conditions are $P$: $\Lcal(B^\dagger B)\ge0$, $Q$: $\Lcal(BB^\dagger)\ge0$, and $G$: $\Lcal(A^\dagger A)\ge0$ for all coefficients.\cite{AyersDavidson2007,Zhao2004} With the contraction, $G$ applied to $A-\Lcal(A)\widehat N/N$ gives the variance form, which is the original nonlinear form of the $G$ condition of Garrod and Percus, with the product of 1-RDMs subtracted,\cite{Garrod1964}
\begin{equation}
 \Lcal(A^\dagger A)-|\Lcal(A)|^2\ge0.
 \label{eq:centered}
\end{equation}
For the number operator $n_c$ of a normalized orbital, $n_c^2=n_c$, so its occupation $\lambda$ satisfies $\lambda(1-\lambda)\ge0$. Thus $G$ gives $0\preceq\gamma\preceq I$, Coleman's condition for ensemble representability of the 1-RDM.\cite{Coleman1963}

The same variance form controls the density covariance
\begin{equation}
 \mathrm{Cov}(x,y)=n_2(x,y)+\rho(x)\delta(x-y)-\rho(x)\rho(y).
 \label{eq:kernel}
\end{equation}
It is positive semidefinite if its quadratic form is nonnegative for every real bounded function $f$,
\begin{subequations}\label{eq:covidentity}
\begin{align}
 C_f&:=\iint f(x)\,\mathrm{Cov}(x,y)\,f(y)\dd x\dd y\\
 &=\int f^2\rho+\iint f(x)f(y)\,n_2(x,y)\dd x\dd y-\Bigl(\int f\rho\Bigr)^2.
\end{align}
\end{subequations}
For a genuine state $C_f$ is the variance of $\int f(x)\hat\psi^\dagger(x)\hat\psi(x)\dd x$, the count weighted by $f$, and is nonnegative. To evaluate it from the orbital moments, write the matrices
\begin{subequations}\label{eq:FTmatrices}
\begin{gather}
 F_{pr}=\int\phi_p^*f\phi_r,\qquad
 T_{pr}=\int\phi_p^*f^2\phi_r,\\
 \dop F=\sum_{pr}F_{pr}a_p^\dagger a_r .
\end{gather}
\end{subequations}
Normal ordering gives $\Lcal\bigl((\dop F)^2\bigr)=\iint ff\,n_2+\gfun{F^2}$, while $\int f^2\rho=\gfun T$. Hence
\begin{equation}
 C_f=\underbrace{\Lcal\bigl((\dop F)^2\bigr)-(\gfun F)^2}_{\text{variance of }\dop F}
     +\underbrace{\gfun{T-F^2}}_{\text{leakage}}.
 \label{eq:covsplit}
\end{equation}
The first term is the variance of the one-body operator $\dop F$ and is nonnegative by Eq.~\eqref{eq:centered}. The second term is nonnegative because $\gamma\succeq0$ and $T-F^2\succeq0$: in terms of the projector $\Pi$ onto the orbital space, $T=\Pi f^2\Pi$ and $F^2=\Pi f\Pi f\Pi$, so that $T-F^2=\Pi f(1-\Pi)f\Pi$, the part of $f$ that leads out of the orbital space and back. It vanishes when $f$ maps the orbital space into itself. Hence $G$ implies $\mathrm{Cov}\succeq0$. Nakata \emph{et al.} obtained this relation by writing the $G$ condition with field operators and noted that $G$ does not enforce the decay of density correlations between separated fragments;\cite{Nakata2002} Eq.~\eqref{eq:covsplit} is its finite-basis form. For $f=\chi_\Omega$, $F=S^\Omega$ and $T=S^\Omega$, and Eq.~\eqref{eq:covsplit} becomes
\begin{equation}
 v_\Omega=\Var_{\Lcal}(\dop{S^\Omega})
          +\gfun{S^\Omega-(S^\Omega)^2}.
 \label{eq:regionalcorrection}
\end{equation}
The same identity is stated in Sec.~II~B of Part I. The $G$ condition thus gives $v_\Omega\ge0$, not the integer bound $\theta(1-\theta)$ of Eq.~\eqref{eq:variancebound}. A table in the supplementary material summarizes the distinctions.

\section{A Gaussian obstruction that survives 3-positivity}
\label{sec:three}
In chemical terms, this section shows that moment data in which each of two orbital groups carries exactly $5/2$ electrons without fluctuation, a fixed fractional charge, pass every positivity test up to degree three, including $T_1$, $T_2$, and $T_2'$, so that 3-positivity does not imply the integer-count conditions. Placed in Gaussian orbitals, these data give a smooth pair density that a half-space count rejects.

A violation of the orbital diagonal conditions can be carried into real space. The diagonal pair occupations $u_i=\gamma_{ii}$ and $d_{ij}=2\Gamma_{ij;ij}$ of a representable 2-RDM lie in the Slater hull,\cite{Kummer1967,AyersDavidson2007} the orbital counterpart of the occupation polytope, in which each orbital holds at most one electron. If the orbitals do not overlap in space and each occupies its own cell, Eq.~\eqref{eq:pairmap} reduces to $n_2(x,y)=\sum_{p\ne q}d_{pq}|\phi_p(x)|^2|\phi_q(y)|^2$, as in Pistol's localized constructions,\cite{Pistol2004} the cell data are exactly $m=u$ and $D=d$, and $n_2$ is representable exactly when $(u,d)$ lies in the Slater hull. Gaussian orbitals always overlap a little, but placing them far apart makes the overlaps exponentially small, so that a strict violation survives at a finite separation while every positivity condition, which involves only the unchanged coefficients, is preserved (general statement and proof in the supplementary material). This section carries this out for a 3-positive input.
\subsection{The input}
Take $N=5$ electrons of one spin in ten orbitals split into two groups $A=\{1,\ldots,5\}$ and $B=\{6,\ldots,10\}$, and the 1- and 2-RDM
\begin{subequations}\label{eq:fiveinput}
\begin{align}
 \gamma&=\tfrac12 I_{10},\\
 \Gamma_{ij;kl}&=\tfrac12d_{ij}(\delta_{ik}\delta_{jl}-\delta_{il}\delta_{jk}),\\
 d_{ij}&=\begin{cases}
 3/16,&i\ne j\text{ in the same group},\\
 1/4,&i,j\text{ in different groups},\\
 0,&i=j.
 \end{cases}
\end{align}
\end{subequations}
Each row of $d$ sums to $4(3/16)+5(1/4)=2=(N-1)\gamma_{ii}$, so the contraction holds. These are the low-order moments of a functional in which each group holds exactly $5/2$ electrons, defined for products of up to three creation and up to three annihilation operators (supplementary material); they are the balanced degree-three moments of a separate manuscript.\cite{NakataDraft} The group number $\widehat K_A=\sum_{i\in A}n_i$ has
\begin{equation}
 \Lcal(\widehat K_A)=\tfrac52,\qquad
 \Lcal(\widehat K_A^2)-\Lcal(\widehat K_A)^2=0,
 \label{eq:orbitalzero}
\end{equation}
so $\Lcal[(\widehat K_A-2)(\widehat K_A-3)]=-1/4$, and the diagonal data lie outside the Slater hull.

\subsection{Full positivity, including mixed lower-degree terms}
By \emph{3-positivity} we mean $\Lcal(W^\dagger W)\ge0$ for every $W$ in the span $\mathscr W_3$ of all products of at most three creation and annihilation operators, including the identity:
\begin{equation}
 \Lcal(W^\dagger W)\ge0\qquad(W\in\mathscr W_3).
 \label{eq:full3}
\end{equation}
This is the third level of the positivity hierarchy of Erdahl and Jin,\cite{ErdahlJin2000,MazziottiErdahl2001} with the mixed lower-degree terms included.

\begin{proposition}[3-positivity of the input]
\label{prop:three}
The input satisfies Eq.~\eqref{eq:full3} and the fixed-$N$ contraction identities through body order three. Its two-body part satisfies $P,Q,G$,\cite{Garrod1964} $T_1,T_2$,\cite{Erdahl1978} and $T_2'$.
\end{proposition}
The analytic proof is given in the supplementary material and is checked there in exact rational arithmetic on all 1351 normal-ordered monomials and all 1161 local-charge Gram blocks. The violated inequality $(\widehat K_A-2)(\widehat K_A-3)\ge0$ is quadratic in the occupations of five orbitals at once; positivity of products of at most three operators does not enforce it.

\subsection{A spatial realization in ten Gaussian-type orbitals}
\label{sec:gaussian}
\begin{figure}[tbp]
 \centering
 \includegraphics[width=\columnwidth]{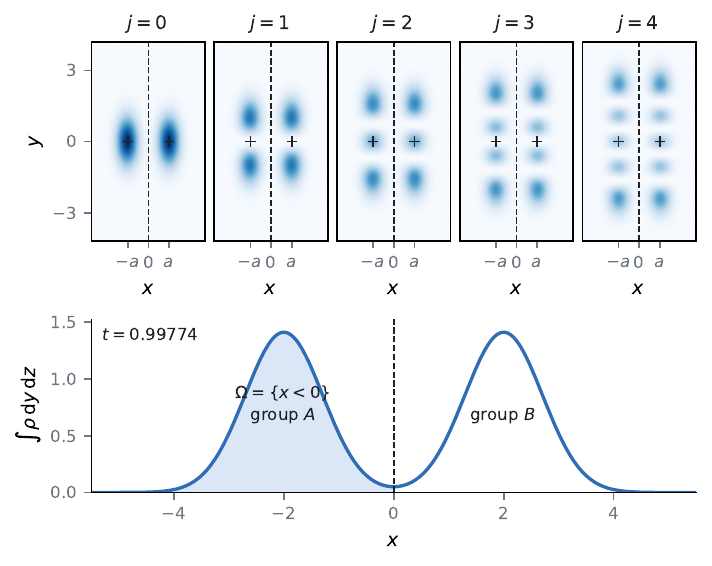}
 \caption{The ten Gaussian-type orbitals of the two-center construction at $a=2$, in units of the Gaussian width. Top: all ten orbitals, shown in pairs; each panel shows the two orbitals $\phi_{L,j}$ (group $A$, at $x=-a$) and $\phi_{R,j}$ (group $B$, at $x=+a$) with the same transverse shape $h_j(y)$, $j=0,\ldots,4$, after symmetric orthogonalization of the pair, as the density $|\phi_{L,j}|^2+|\phi_{R,j}|^2$ in the plane $z=0$; crosses mark the two centers, the dashed line the boundary $x=0$ of $\Omega$, and the $j+1$ spots along $y$ come from the $j$ nodes of $h_j$. Orbitals with different $j$ are orthogonal by Eq.~(\ref{eq:orbitals}c), so that $\iiint\phi_{s,j}\phi_{s',j'}\dd x\dd y\dd z=0$ for $j\ne j'$ and $s,s'\in\{L,R\}$; within a pair, the symmetric orthogonalization removes the overlap $e^{-a^2}$ of $g_L$ and $g_R$. Bottom: the one-electron density $\rho=\tfrac12\sum_i|\phi_i|^2$ integrated over $y$ and $z$; the half space $\Omega=\{x<0\}$ (shaded) contains a fraction $t$ of each group-$A$ orbital, Eq.~(\ref{eq:t}b).}
 \label{fig:gaussianorbitals}
\end{figure}

The input is placed in real space as described at the beginning of this section; a two-center construction gives closed formulas. With
\begin{subequations}\label{eq:orbitals}
\begin{gather}
 g_{L,R}(x)=\pi^{-1/4}e^{-(x\pm a)^2/2},\\
 h_j(y)=\frac{H_j(y)\,e^{-y^2/2}}{\sqrt{2^jj!\sqrt\pi}},\\
 \int h_j(y)h_{j'}(y)\dd y=\delta_{jj'},
\end{gather}
\end{subequations}
where $H_j$ is the Hermite polynomial of degree $j$, place the five Gaussian-type functions $L_j=g_L(x)h_j(y)h_0(z)$ at $x=-a$ and the five $R_j=g_R(x)h_j(y)h_0(z)$ at $x=+a$, $j=0,\ldots,4$, all with the same spin. In the order $(L_0,\ldots,L_4,R_0,\ldots,R_4)$, with $s=e^{-a^2}$ and $u=\tfrac12(1+\operatorname{erf}a)$, their overlap matrices over all space and over the half space $\Omega=\{x<0\}$ are
\begin{equation}
 O=\begin{pmatrix}I_5&sI_5\\ sI_5&I_5\end{pmatrix},\qquad
 J=\begin{pmatrix}uI_5&\tfrac s2I_5\\ \tfrac s2I_5&(1-u)I_5\end{pmatrix}.
 \label{eq:OJmain}
\end{equation}
Functions with different $j$ do not overlap by Eq.~(\ref{eq:orbitals}c), so only $L_j$ and $R_j$ overlap, and every matrix is a $2\times2$ matrix times $I_5$. Symmetric orthogonalization with $O^{-1/2}$ gives ten orthonormal orbitals $\phi_{L,j}$ and $\phi_{R,j}$; group $A$ is the left center and group $B$ the right. Their half-space overlap matrix $S^\Omega=O^{-1/2}JO^{-1/2}$ is exactly diagonal,
\begin{subequations}\label{eq:t}
\begin{align}
 S^\Omega&=\diag(tI_5,(1-t)I_5),\\
 t=t(a)&=\frac12\left(1+
 \frac{\operatorname{erf}(a)}{\sqrt{1-e^{-2a^2}}}\right),
\end{align}
\end{subequations}
at every finite $a$: each group-$A$ orbital has the weight $t$ in $\Omega$ and each group-$B$ orbital the weight $1-t$ (calculation in the supplementary material). At $a=2$, $s=0.0183$, $u=0.99766$, and $t=0.99774$. Figure~\ref{fig:gaussianorbitals} shows the ten orbitals and the half space.

\subsection{Nonrepresentability despite 3-positivity}
Placed in real space, the two orbital groups with exactly $5/2$ electrons each still pass every positivity test up to degree three, but counting the electrons on one side of a plane shows too little fluctuation for an integer count. This is the real-space form of the failure in which a molecule dissociates into fractionally charged atoms.
\begin{theorem}[A three-positive Gaussian counterexample]
\label{thm:gaussian}
Use the coefficients of Eq.~\eqref{eq:fiveinput} in the ten orthonormal Gaussian-type orbitals above, with $a=2$. They are 3-positive (Proposition~\ref{prop:three}), and their 1-RDM is pure-state representable. Their pair density is smooth, nonnegative, symmetric, normalized to 20, and has positive semidefinite density covariance. Nevertheless, it is not the pair density of any five-electron probability distribution.
\end{theorem}
\begin{proof}[Proof sketch]
The pair density is a positive combination of squared antisymmetrized orbital products, which gives the regularity and nonnegativity; $\gamma=I_{10}/2$ is the 1-RDM of $(|12345\rangle+|6\,7\,8\,9\,10\rangle)/\sqrt2$; and the density covariance is positive semidefinite by Eq.~\eqref{eq:covsplit}, since the input satisfies $G$ (Proposition~\ref{prop:three}) and $\gamma=I_{10}/2\succeq0$. Equation~\eqref{eq:t} in the contractions gives
\begin{equation}
 \mu_\Omega=\frac52,\quad
 v_\Omega=5t(1-t),\quad
 I_{\Omega,2}=v_\Omega-\frac14.
 \label{eq:gaussianmoments}
\end{equation}
At $a=2$, $v_\Omega=0.01125\ldots$ and $I_{\Omega,2}=-0.23875\ldots$; a rational bound, $I_{\Omega,2}<-1801/8000$, makes the sign independent of the decimals. An actual count $N_\Omega$ takes integer values, and no integer lies strictly between 2 and 3, so $(N_\Omega-2)(N_\Omega-3)\ge0$ for every value and its average $I_{\Omega,2}$ is nonnegative for every five-electron distribution. With mean $5/2$,
\begin{equation}
 \bigl\langle(N_\Omega-2)(N_\Omega-3)\bigr\rangle=v_\Omega+\bigl(\tfrac52-2\bigr)\bigl(\tfrac52-3\bigr)=v_\Omega-\tfrac14,
 \label{eq:countexpand}
\end{equation}
so a count fixed at $5/2$ would give $\tfrac12\cdot(-\tfrac12)=-\tfrac14$, and the candidate, with $v_\Omega=0.011$, gives a negative value. Hence no five-electron distribution has this pair density. The complete proof is given in the supplementary material.
\end{proof}

The contrast can be stated without a matrix qualification:
\begin{equation}
 \boxed{\begin{gathered}
 n_2\ge0\ \text{and smooth},\qquad \mathrm{Cov}\succeq0,\\
 \gamma\ \text{pure-state representable},\\
 \mu_\Omega=5/2,\qquad v_\Omega<1/4.
 \end{gathered}}
 \label{eq:centralcontrast}
\end{equation}
Thus even with 3-positivity, the spatial two-body data do not support the elementary probability interpretation of an integer electron count. The failure concerns the pair density itself, and the pair density is all that the count inequality needs: changing the basis, choosing another 2-RDM, or adding empty orbitals toward a complete set, which preserves 3-positivity (supplementary material), cannot repair the \emph{same} pair density. Read in the other direction, this is a test of the 2-RDM itself. Placing the orbitals of an $N$-representable 2-RDM onto any orthonormal set of spatial orbitals, including orbitals with disjoint compact supports, of which any number can be constructed, maps its $N$-electron state to an $N$-electron state and therefore gives a representable pair density. A single orbital realization whose pair density violates an integer count thus shows that the 2-RDM is not $N$-representable. A separate manuscript shows, with an exactly solvable counting operator, that the $(2,q)$-positivity hierarchy needs a depth that grows linearly with $N$ to exclude such zero-variance inputs,\cite{NakataDraft} whereas the count inequality on the pair density excludes them at once. Equation~\eqref{eq:regionalcorrection} identifies the origin of the small positive $v_\Omega$: the orbital-space count $t\widehat K_A+(1-t)\widehat K_B$ has zero variance in the input, and the entire real-space variance is the leakage term $\gfun{S^\Omega-(S^\Omega)^2}=5t(1-t)$ from the orbital tails. Positivity sees this correction but does not enforce the integer lower bound. Figure~\ref{fig:variance} shows its dependence on the separation.

\subsection{Interpretation: fractionally charged fragments at dissociation}
\label{sec:dissociation}
\begin{figure}[tbp]
 \centering
 \includegraphics[width=\columnwidth]{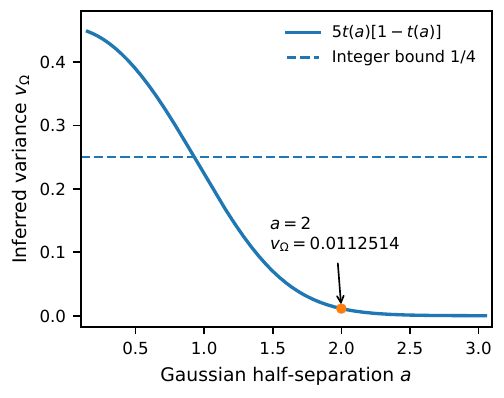}
 \caption{Inferred half-space variance $v_\Omega(a)=5t(a)[1-t(a)]$ for the unchanged 3-positive input, versus the half-separation $a$ of the two centers in units of the Gaussian width. The dashed line is the integer-count lower bound $1/4$ at mean $5/2$. The marked point is the finite construction $a=2$. The curve is sampled from the analytic expression; the counterexample is proved by a rational bound, not by the plot.}
 \label{fig:variance}
\end{figure}

The two-center construction has a direct chemical reading. The two groups of five orbitals are localized on the centers at $\mp a$, so the input describes a stretched symmetric system of five same-spin electrons, and $\Omega$ contains one fragment. For any five-electron state, the number of electrons in $\Omega$ is an integer random variable. With mean $5/2$ it must take at least two values, and $v_\Omega\ge1/4$; the bound is attained by the equal mixture of the fragment configurations $(2,3)$ and $(3,2)$, which is exactly the repaired law of Sec.~\ref{sec:gaussianrepair}. The candidate instead has $v_\Omega=5t(1-t)\to0$ as $a\to\infty$ (Fig.~\ref{fig:variance}), and $I_{\Omega,2}\to-1/4$. In the dissociation limit it therefore describes two fragments each carrying $5/2$ electrons with vanishing fluctuation: fractionally charged fragments.

Fractional charges at dissociation are the known failure of approximate theories recalled in Sec.~\ref{sec:intro}. Theorem~\ref{thm:gaussian} shows that the corresponding real-space signature can survive all 3-positivity tests, including $T_1$, $T_2$, and $T_2'$. The regional count inequality, and more generally the occupation LP, detects it from the pair density alone, without a Hamiltonian or fragment energies. We do not claim that a variational calculation for a particular molecule selects this input; the theorem shows only that degree-three positivity does not exclude it. A molecular relaxation with the same signature was observed in Part I:\cite{NakataPartI} for N$_2^+$ at 4~\AA{} in a minimal basis (Sec.~V and Table~IV of Part I), the solutions with the $P$, $Q$, and $G$ conditions and with the additional $T_1$ and $T_2$ conditions keep the half-integer mean population of each atomic region, but their count variance lies far below the bound $1/4$, and one region cut restores the bound. That calculation does not impose $T_2'$, and its 2-RDM differs from the input constructed here, so it is an instance of the same mechanism rather than a realization of Theorem~\ref{thm:gaussian}. The spin-resolved counterpart, an integer fragment charge with a fractional fragment spin, is the example of Eq.~\eqref{eq:spinexample}.

\subsection{The same Gaussian violation as an exact finite repair problem}
\label{sec:gaussianrepair}
The example also connects the exclusion theorem to the constructive LP. For the two cells $\Omega$ and $\Omega^c$, with $v=5t(1-t)<1/4$,
\begin{equation}
 m=(5/2,5/2),\qquad
 D=\begin{pmatrix}15/4+v&25/4-v\\25/4-v&15/4+v\end{pmatrix}.
 \label{eq:gaussianD}
\end{equation}
At these means the smallest allowed $\langle N_\Omega(N_\Omega-1)\rangle$ is 4, so the nearest feasible table and the two-cell distance are
\begin{equation}
 \widetilde D=\begin{pmatrix}4&6\\6&4\end{pmatrix},\qquad
 \delta_{\{\Omega,\Omega^c\}}=1-4v,
 \label{eq:gaussianrepair}
\end{equation}
realized by the equal mixture of $(2,3)$ and $(3,2)$. At $a=2$, $\delta=0.95499\ldots$. An upper bound needs no six-dimensional integral: the equal mixture of five-electron determinants with group populations $(2,3)$ or $(3,2)$ and uniformly chosen occupied orbitals in the same ten orbitals has exactly the candidate's $\rho$ and a pair density within $L^1$ distance 1. Hence
\begin{equation}
 0.95499\ldots\le d_\rho(n_2)\le1\qquad(a=2),
 \label{eq:gaussianupper}
\end{equation}
and the distances of this family tend to one as $a\to\infty$ (derivation and exact check in the supplementary material).

\section{Application to variational 2-RDM solutions}
\label{sec:molecules}
The finite diagnosis applies to any candidate pair density, including that of a relaxed variational 2-RDM. In a finite orbital basis the cell data are the contractions of Eq.~\eqref{eq:pullback}, so the occupation LP needs only $(\gamma,\Gamma)$. Two of its outputs are used here. For real-space cells the optimum $\delta_{\mathcal P}$ is a lower bound on the fixed-$\rho$ distance of the relaxed pair density (Theorem~\ref{thm:distance}). The values reported below are numerical estimates for the computed SDP solutions, whose accuracy is documented in the supplementary material; they are not certified bounds in the sense of Sec.~\ref{sec:constructive}~C, which would require propagating the solution errors into the dual value. For spin-summed channels the LP bounds $d_\rho(n_2)$, and for jointly spin-resolved channels it bounds $d^{\mathrm{spin}}$ of Proposition~\ref{prop:spin}, which can be larger. A dual-feasible solution with a positive objective value supplies an inequality that the relaxation violates, that every $N$-electron state (with the fixed $N_\uparrow$, $N_\downarrow$ in the spin-resolved case) satisfies, and that is affine in the 2-RDM by Eq.~\eqref{eq:fixedcut}; added to the semidefinite program (SDP), it acts as a cutting plane. Each such cut is the expectation value of a two-body operator that is positive semidefinite on the particle-number sector used in the optimization, with fixed $N_\uparrow$ and $N_\downarrow$ for the jointly spin-resolved cuts, and in the dual SDP it enters with a nonnegative multiplier; in the weighted sum-of-squares language of Rubin, Low, and DePrince,\cite{Rubin2026} the dual lower-bound certificate thus acquires positive terms besides the Gram terms of the imposed positivity conditions. The primal lift of Eq.~\eqref{eq:densitylift} leaves the orbital space, so it does not provide a repaired 2-RDM and is not used. All partitions are real-space partitions, and no Hartree--Fock (HF) calculation enters: neither the orbital basis, nor the placement of the cells, nor the construction of the cuts uses HF orbitals or densities. The systems include the dissociations of Part I,\cite{NakataPartI} whose region-cut solutions are compared with the present ones, and small polyatomic systems in which a single region per bond is not sufficient.

The applications are the following.
\begin{itemize}
 \item Distance lower bounds (Sec.~\ref{sec:distancebounds}): the LP optima on nested real-space partitions bound the distance of each relaxed pair density from every representable one, and their meaning is illustrated for stretched N$_2^+$.
 \item Partitions chosen by the LP optimum (Sec.~\ref{sec:comparepartI}): the dual witnesses are added as cuts on partitions fixed by the geometry or chosen by the size of the LP optimum, and the results are compared with the region cuts of Part I, for diatomic molecules and for polyatomic chains.
 \item Fractional electron numbers (Sec.~\ref{sec:fractional}): Fock-space relaxations with the integer-count condition for atoms between $Z-1$ and $Z+1$ electrons, compared with the exact piecewise-linear energy.
 \item Remaining errors (Sec.~\ref{sec:remaining}): the errors that remain after every tested partition passes.
\end{itemize}

\subsection{Computational details}
\label{sec:compdetails}
The molecular calculations use the STO-6G basis; the atomic calculations of Sec.~\ref{sec:fractional} also use 6-31G. The orbital basis is the canonically orthonormalized atomic-orbital basis $X=VW^{-1/2}$, with $S=VWV^{\mathsf T}$ the atomic-orbital overlap matrix of the molecule, or of each atom for the separated atoms; no self-consistent-field calculation is performed, and the relaxations are invariant to this choice of orthonormal basis. The full configuration interaction (FCI) references for the lowest state of the target spin were computed in an HF orbital basis spanning the same atomic-orbital space; their 1- and 2-RDMs are rotated to the present basis, reproduce the FCI energies to $10^{-8}$~$E_\mathrm h$, and are used only to evaluate errors and to check that every cut holds for a physical state. The relaxations impose $P$, $Q$, $G$, or $P$, $Q$, $G$, $T_1$, $T_2$, together with the spin-resolved traces, the contractions, and, for the molecules, $\langle\hat S_-\hat S_+\rangle=0$, and are solved with SDPA 7.4.3;\cite{Yamashita2003} this is the equality variant of Part I without its additional $\langle\hat L_z^2\rangle$ condition. The diatomic molecules are N$_2^+$ at 4~\AA, NO$^+$ at 2.5--6~\AA, CN$^-$ at 4 and 6~\AA, BeO$^+$ at 4~\AA, and the separated-atom limits of N$_2^+$, NO$^+$, and CO$^+$, built from atomic orbitals with all interatomic integrals omitted; the polyatomic systems are linear H$_4^+$ and H$_6^+$ chains and symmetrically stretched H$_2$O. The baseline SDP energies reproduce those of Part I within the solver accuracy. For Sec.~\ref{sec:fractional}, the atoms Be to O are treated in the STO-6G and 6-31G bases with all electrons, between $Z-1$ and $Z+1$ electrons in steps of 0.1; the added electron of $N=k+\theta$ is assigned to one spin, so that one spin mean stays integer. Their FCI references are computed directly in the same orbital basis, as the lowest of several roots. For the separated-atom comparison the atomic relaxations are evaluated on a grid of spin means of step $1/8$, and for NO$^+$ in 6-31G the 1s orbital of each atom, the natural orbital of largest occupation of the neutral-atom FCI, is frozen. The region-cut solutions of Part I are not recomputed: their stored primal vectors are reanalyzed, and their energies are recomputed from these vectors, for the comparison in Table~\ref{tab:compare}.

For a diatomic molecule on the $z$ axis, with atom $A$ at the origin and atom $B$ at $z=R$, the cells are slabs between planes perpendicular to the bond. The basic plane is the Voronoi split plane $z_{\mathrm V}=Rr_A/(r_A+r_B)$, where the boundary between the atom cells weighted by the Bragg radii $r_A$, $r_B$ crosses the bond axis. The fixed partitions are the two cells at $z_{\mathrm V}$, the four slabs bounded at the nuclei and at $z_{\mathrm V}$, the two cells at $z_{\mathrm V}$ split further at $x=0$, and nested partitions with 2, 4, and 8 cells used for refinement. For the separated atoms each atom has its own copy of space, and the cells are the region of one atom, its half spaces through the nucleus normal to $x$ or $z$, and balls of radius 0.6, 1.2, and 2.0 bohr about the nucleus. For the polyatomic systems every point is assigned to the atom $A$ that minimizes $|\bm r-\bm R_A|/r_A$; unions of these atom cells define fragment regions. Cells bounded by planes perpendicular to the coordinate axes are integrated exactly with error functions, the atom-centered cells of the separated atoms by product Gauss quadrature with no quadrature point on a cell boundary, and the curved cells of H$_2$O on a DFT grid; the accuracy of all cell matrices is documented in the supplementary material. Each cell is a channel for the spin-summed count or, jointly spin-resolved (Proposition~\ref{prop:spin}), two channels for the $\alpha$ and $\beta$ counts; all integer configurations are enumerated.

The dual problem of Eq.~\eqref{eq:LPpair} is solved in floating point; each cut is shifted so that it holds on every configuration and is checked on the FCI 2-RDM (supplementary material). In each iteration one cut is added for every partition with $\delta_{\mathcal P}>10^{-5}$, all previous cuts are kept, and the SDP is solved again until every partition has $\delta_{\mathcal P}\le10^{-5}$. The $P,Q,G,T_1,T_2$ separations start from the converged cuts of the corresponding $P,Q,G$ run.

Because every partition yields valid cuts, the partition can also be chosen anew from each SDP solution. In the plane search, planes on a grid of spacing $\adaptStepnop$~bohr from $3$~bohr before atom $A$ to $3$~bohr beyond atom $B$, including $z_{\mathrm V}$, are scanned for the single plane whose jointly spin-resolved two-cell LP has the largest optimum and for the plane whose insertion into the current refined slab partition gives the largest spin-summed optimum, up to eight cells; the witnesses with $\delta>10^{-5}$ are added and the search is repeated on the new solution. As a chemically motivated alternative, the partitions are regenerated at the minimum of the density of the current solution on the bond axis, its bond critical point. The search is a heuristic over a nonconvex family of partitions: a violation that it finds is valid, but the largest violation need not be found. The supplementary material lists the solver status, primal--dual gaps, equality residuals, and eigenvalue residuals of all solutions, and the partitions and spin modes of all runs.

\subsection{Distance lower bounds of relaxed pair densities}
\label{sec:distancebounds}
For FCI every LP optimum vanishes to numerical precision, and for the relaxed solutions the optima of nested partitions do not decrease under refinement, as Theorem~\ref{thm:distance} requires (supplementary material). For NO$^+$ a violation of $\refDeltaSepFineNOp$ appears only on the eight-cell partition after the two- and four-cell cuts: a pass at a coarse partition does not exclude a violation under refinement.

The meaning of these numbers is easiest to see for N$_2^+$ at 4~\AA, using the nested slabs with up to eight cells (Table~\ref{tab:deltameaning}). The $P,Q,G$ solution gives each atom the correct mean count $13/2$, but an inferred variance $v=\rvPqgnTwoFourZero$ on the two cells at the bond midpoint: far below the bound $1/4$ that every distribution of integer electron numbers with this mean obeys. An integer count with mean $13/2$ has a variance of at least $1/4$, reached when six and seven electrons occur with equal probability. To reach it, each of the four entries of the two-cell pair table must change by $1/4-v$, so that $\delta=1-4v=\ePdeltaPqgnTwoFourZero$ of the total ordered-pair mass $\ePmassnTwoFourZero$. This is a lower bound on the distance of the relaxed pair density from every representable one. The value is the same for two, four, and eight slabs, because the violation lies entirely in the statistics of the atom counts, which the two cells at the bond midpoint already resolve. After the cuts, $\delta<10^{-5}$ for all these partitions, and the variance is $v=\rvCutsnTwoFourZero$, at the bound $1/4$ within $\rvCutsGapnTwoFourZero$, the accuracy of the solver (FCI: $v=\rvFcinTwoFourZero$); within this accuracy the cell data are those of a distribution in which six and seven electrons occur on each atom with equal probability. This does not prove that the pair density is representable. Theorem~\ref{thm:distance} gives only $d_\rho\le\delta+e_{\mathcal P}$, and $e_{\mathcal P}=\ePslabCutsnTwoFourZero$ for the eight slabs, so the distance is merely bounded by that value. The remaining error of $\cmpFixednTwoFourZero$~m$E_\mathrm h$ is therefore not accompanied by any violation that these slabs detect.

\begin{table}[tbp]
\caption{Meaning of the LP optimum for N$_2^+$ at 4~\AA{} (nested slabs with up to eight cells, spin-summed).}
\label{tab:deltameaning}
\begin{ruledtabular}
\begin{tabular}{p{0.2\columnwidth}p{0.12\columnwidth}p{0.58\columnwidth}}
Solution & $\delta$ & Meaning\\
\hline
$P,Q,G$ & $\ePdeltaPqgnTwoFourZero$ & Atom-count fluctuation too small for integer electrons ($v=\rvPqgnTwoFourZero$); distance at least $\ePdeltaPqgnTwoFourZero$ (of $\ePmassnTwoFourZero$)\\[3pt]
$P,Q,G$ + cuts & $<10^{-5}$ & No violation visible on these slabs; $v=\rvCutsnTwoFourZero$; representability not established; distance at most about $e_{\mathcal P}\approx\ePslabCutsnTwoFourZero$ (numerical estimate)\\
\end{tabular}
\end{ruledtabular}
\end{table}

The upper bound of Theorem~\ref{thm:distance} is far less informative: the coarse-graining error $e_{\mathcal P}$ stays near $\ePslabnTwoFourZero$ for slabs and decreases only on partitions much finer than those on which the LP can be solved ($\ePfinenTwoFourZero$ with $\ePfineCellsnTwoFourZero$ cells). A known representable pair density gives a much tighter bound. After the cuts, the pair densities of NO$^+$ and BeO$^+$ at 4~\AA{} lie within $\fciDistCutsnopFourZero$ and $\fciDistCutsbeoFourZero$ of the FCI ones, whereas that of N$_2^+$ is still $\fciDistCutsnTwoFourZero$ away. Because the FCI one-electron density differs from $\rho$, this is not $d_\rho$; moving the FCI electrons to $\rho$ costs at most $2(N-1)\|\rho-\rho_{\mathrm{FCI}}\|_1$ (supplementary material), which gives the numerical estimates $d_\rho\lesssim\fixRhoBoundCutsnopFourZero$ for NO$^+$ and $\fixRhoBoundCutsbeoFourZero$ for BeO$^+$, compared with about $\ePslabCutsnopFourZero$ and $\ePslabCutsbeoFourZero$ from the eight slabs. The relaxed pair density of NO$^+$ with cuts is thus close to the representable set at the same $\rho$; for N$_2^+$ no such conclusion follows.

\subsection{Comparison with Part I: partitions chosen by the LP optimum}
\label{sec:comparepartI}
% generated by calc/make_tex.py from calc/derived; do not edit
\begin{table*}[tbp]
\caption{Errors $E-E_{\mathrm{FCI}}$ (m$E_\mathrm h$, primal objectives) of minimal-basis (STO-6G) 2-RDM relaxations
with real-space cuts, compared with the stored region-cut solutions of Part I (reanalyzed, energies recomputed from the
stored primal vectors). Part I: one region bounded at the bond critical point of the Hartree--Fock density (atom
regions at $R=\infty$). Present work, no Hartree--Fock input: Voronoi, cuts of the two-cell partition at the Voronoi
split plane, the four slabs bounded at the nuclei and that plane, and that plane combined with $x=0$ (at $R=\infty$:
the real-space atom cells, half spaces through each nucleus, and balls of radius 0.6, 1.2, and 2.0 bohr about each
nucleus); BCP: two- to eight-cell partitions regenerated at the density minimum of each solution on the bond axis;
search: a plane chosen anew from each solution by the largest LP optimum among the scanned planes, with insertion into a refined slab partition
(Sec.~\ref{sec:molecules}). With $T_1$ and $T_2$, the separation starts from the converged $P,Q,G$ cuts of the search. --: not applicable (BCP at $R=\infty$, where no bond axis separates the atoms). $^\dagger$SDPA phase pFEAS with $|E_P-E_D|>0.05$~m$E_\mathrm h$ for the final solution (supplementary material). $^\ddagger$At $R=\infty$ no plane is searched, and the search column repeats the Voronoi partitions.}
\label{tab:compare}
\begin{ruledtabular}
\begin{tabular}{lrrrrrrrr}
 & \multicolumn{5}{c}{$P,Q,G$} & \multicolumn{3}{c}{$P,Q,G,T_1,T_2$}\\
System & None & Part I & Voronoi & BCP & Search & None & Part I & Search\\
\hline
N$_2^+$, 4~\AA & $-137.98$ & $-22.05$ & $-22.05$ & $-22.05$ & $-22.05$ & $-124.79$ & $-1.626$ & $-1.625$\\
NO$^+$, 2.5~\AA & $-72.70$ & $-15.76$ & $-11.01$ & $-10.41$ & $-8.930$ & $-66.39$ & $-11.06$ & $-5.139$\\
NO$^+$, 3.0~\AA & $-87.00$ & $-3.496$ & $-2.103$ & $-2.014$ & $-0.979$$^\dagger$ & $-83.77$ & $-2.685$ & $-0.458$\\
NO$^+$, 3.5~\AA & $-93.92$ & $-0.683$ & $-0.365$ & $-0.368$ & $-0.094$ & $-91.08$ & $-0.560$ & $-0.035$\\
NO$^+$, 4~\AA & $-98.41$ & $-0.146$ & $-0.071$ & $-0.074$ & $-0.014$ & $-95.76$ & $-0.124$ & $-0.003$\\
NO$^+$, 6.0~\AA & $-109.02$ & $-0.002$ & $0.000$ & $-0.001$ & $0.001$ & $-106.45$ & $-0.004$ & $0.001$\\
CN$^-$, 4.0~\AA & $-70.50$ & $-0.218$ & $-0.333$ & $-0.320$ & $-0.084$ & $-67.93$ & $-0.145$ & $-0.032$\\
CN$^-$, 6.0~\AA & $-80.75$ & $-0.005$ & $-0.004$ & $-0.003$ & $-0.004$ & $-78.32$ & $-0.003$ & $-0.001$\\
BeO$^+$, 4.0~\AA & $-32.03$ & $-1.254$ & $-0.696$ & $-1.195$ & $-0.489$ & $-17.51$ & $-0.598$ & $-0.153$\\
N$_2^+$, $R=\infty$ & $-170.42$ & $-21.54$ & $-21.54$ & -- & $-21.54$$^\ddagger$ & $-158.10$ & $-1.431$ & $-1.433$$^\ddagger$\\
NO$^+$, $R=\infty$ & $-130.58$ & $-0.003$ & $-0.001$ & -- & $-0.001$$^\ddagger$ & $-127.99$ & $-0.004$ & $0.001$$^\ddagger$\\
CO$^+$, $R=\infty$ & $-156.40$ & $-19.51$ & $-19.51$ & -- & $-19.51$$^\ddagger$ & $-137.29$ & $-5.020$ & $-5.013$$^\ddagger$\\
\end{tabular}
\end{ruledtabular}
\end{table*}

In Part I the cut candidates were constructed in advance from the HF density and orbitals, and the region was bounded at the bond critical point of the HF density. Here each cut is the dual witness of the current solution, and the partitions are either fixed by the geometry (the Voronoi plane) or chosen by the LP optimum. Table~\ref{tab:compare} compares the two. Where the fragments of Part I carry integer or exactly half-integer mean charges, the present cuts reproduce the region-cut energies of Part I within the solver accuracy without HF input: N$_2^+$ at 4~\AA{} and at the separated-atom limit, NO$^+$ and CN$^-$ at 6~\AA, and the separated atoms of NO$^+$ and CO$^+$. For the separated atoms the half spaces and balls about each nucleus detect the same violation as the atom cells and give no further energy change.

\begin{figure*}[tbp]
 \centering
 \includegraphics[width=\textwidth]{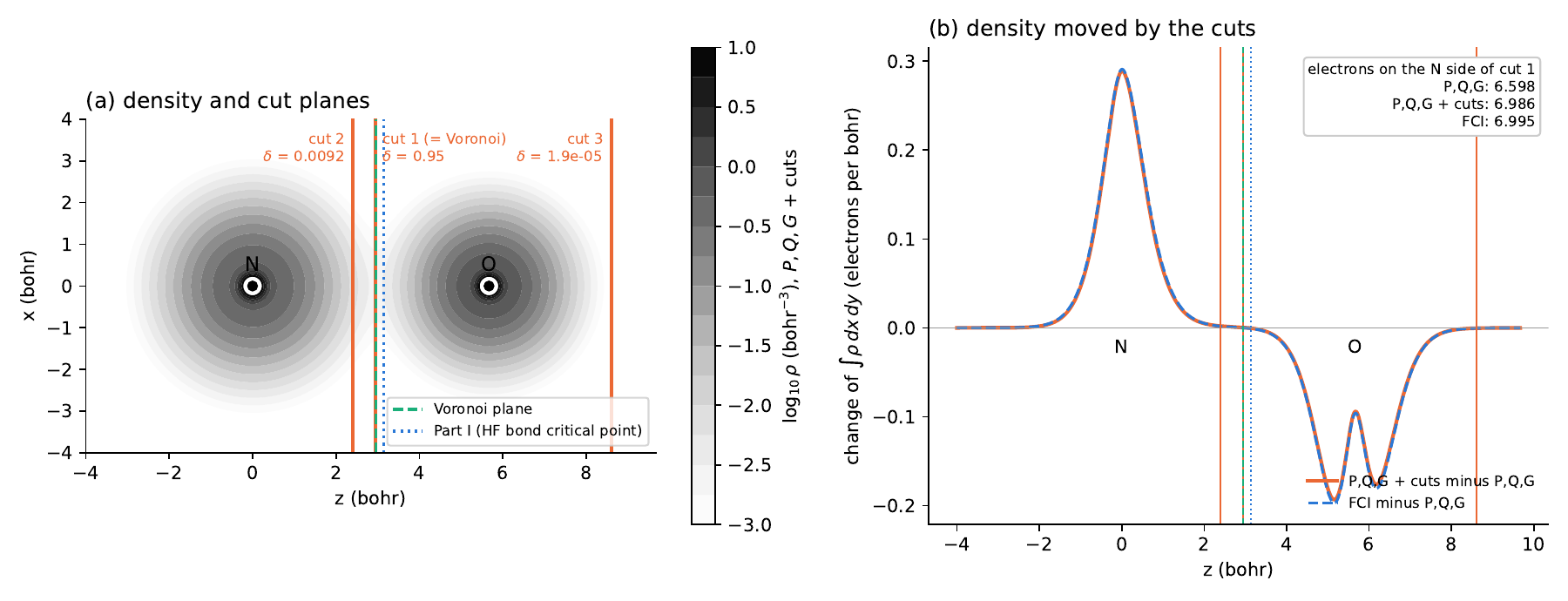}
 \caption{Plane search for NO$^+$ at 3~\AA{} (STO-6G, $P,Q,G$). (a) Electron density of the final solution with cuts in a plane containing the bond axis, with the planes whose witnesses were added (solid; $\delta$ is the LP optimum when the plane was found), the Voronoi plane (dashed), and the Part I region boundary at the bond critical point of the Hartree--Fock density (dotted). (b) Where the cuts move electrons: the number of electrons per unit length along the bond, integrated over each plane perpendicular to it, minus that of the $P,Q,G$ solution without cuts, for the solution with cuts (solid) and for FCI (dashed); positive values mean more electrons than without cuts, so the cuts move electrons from O back to N. The two curves nearly coincide, so the redistribution produced by the cuts is almost the one that leads to FCI: relative to FCI, the solution without cuts has moved $\cutFigExcessPqg$ electron too many from N to O, and with the cuts this excess charge transfer is only $\cutFigExcessCuts$ electron. The inset gives the number of electrons on the N side of the first plane. The third plane carries a violation close to the threshold $10^{-5}$ and changes the energy by about the solver accuracy.}
 \label{fig:cutplanes}
\end{figure*}

At intermediate distances, where a fraction of an electron is transferred, the position of the boundary matters. For NO$^+$ the cuts at the fixed Voronoi plane already give smaller errors than the region cut of Part I, for example $\cmpFixednopThreeZero$ instead of $\cmpPartInopThreeZero$~m$E_\mathrm h$ at 3~\AA, whereas for CN$^-$ at 4~\AA{} they do not ($\cmpFixedcnmFourZero$ instead of $\cmpPartIcnmFourZero$~m$E_\mathrm h$). For NO$^+$ and CN$^-$, regenerating the partitions at the bond critical point of each solution changes these errors only slightly. Figure~\ref{fig:cutplanes} shows the search for NO$^+$ at 3~\AA: the first plane coincides with the Voronoi plane and reduces the excess charge transfer relative to FCI from $\cutFigExcessPqg$ to $\cutFigExcessCuts$ electron in the final solution, and two further planes carry much smaller residual violations. The plane search, which selects the plane by the size of the LP optimum, gives the smallest errors in every such case: $\cmpAdaptnopTwoFive$, $\cmpAdaptnopThreeZero$, $\cmpAdaptnopThreeFive$, and $\cmpAdaptnopFourZero$~m$E_\mathrm h$ for NO$^+$ at 2.5, 3, 3.5, and 4~\AA, compared with $\cmpPartInopTwoFive$, $\cmpPartInopThreeZero$, $\cmpPartInopThreeFive$, and $\cmpPartInopFourZero$~m$E_\mathrm h$ in Part I, and $\cmpAdaptcnmFourZero$ instead of $\cmpPartIcnmFourZero$~m$E_\mathrm h$ for CN$^-$ at 4~\AA. Several of these SDPA solutions end with phase pFEAS even after a second solution with more stable SDPA parameters (supplementary material), with primal--dual gaps up to $\cmpAdaptGapnopThreeZero$~m$E_\mathrm h$ (supplementary material); without confirmed dual feasibility such a gap is an estimate of the accuracy, not a certified interval around the optimum; the differences from Part I at 2.5 and 3~\AA{} exceed these gaps, whereas differences of the size of the gaps are not significant. Each iteration of the search adds both a spin-resolved two-cell witness and a spin-summed witness of the refined slab partition. A control limited to two cells gives $\cmpTwoCellnopTwoFive$ and $\cmpTwoCellnopThreeZero$~m$E_\mathrm h$ for NO$^+$ at 2.5 and 3~\AA, compared with $\cmpAdaptnopTwoFive$ and $\cmpAdaptnopThreeZero$~m$E_\mathrm h$ with the refined slabs, with primal--dual gaps of up to $\cmpAdaptGapnopThreeZero$~m$E_\mathrm h$ (phase pFEAS, so not certified intervals). The joint spin-resolved two-cell witnesses accumulated over the searched planes thus reproduce most of the gain over Part I, which therefore cannot be assigned to the position of the first plane alone; an additional effect of the slab refinement cannot be separated at this accuracy. Part I also imposed an additional $\langle\hat L_z^2\rangle$ condition (Sec.~\ref{sec:compdetails}), so the comparison with Part I mixes the choice of partition with this difference in the constraints; the effect of the partition alone is seen in the Voronoi, BCP, and search columns, which share the same constraints. For NO$^+$ at 4~\AA{} the first plane lies at $z=\adaptPlaneOnenop$~bohr, at the Voronoi plane $z_{\mathrm V}=\zsplitnop$~bohr, and has the effect of the region cut; a second plane at $z=\adaptPlaneTwonop$~bohr then carries a violation $\delta=\adaptDeltaTwonop$, and the iteration ends with $\adaptCutsnopPQG$ cuts; with half the grid spacing the result changes by less than $\adaptStepBoundnop$~m$E_\mathrm h$. The planes with the largest violations need not lie at a bond critical point, and the LP optimum, a lower bound on a distance rather than a yes-or-no test, provides the criterion by which they are found.

With $T_1$ and $T_2$ the converged $P,Q,G$ cuts are added from the start and the plane search is continued; it converged within one or two iterations for every system. Without cuts, $T_1$ and $T_2$ reduce the dissociation error only partially (columns ``None''), for example from $\cmpBasebeoFourZero$ to $\cmpTtBasebeoFourZero$~m$E_\mathrm h$ for BeO$^+$. With the searched planes the errors are again smaller than those of Part I wherever a fraction of an electron is transferred: $\cmpTtnopTwoFive$, $\cmpTtnopThreeZero$, $\cmpTtnopThreeFive$, and $\cmpTtnopFourZero$~m$E_\mathrm h$ for NO$^+$ at 2.5, 3, 3.5, and 4~\AA, compared with $\cmpPartITtnopTwoFive$, $\cmpPartITtnopThreeZero$, $\cmpPartITtnopThreeFive$, and $\cmpPartITtnopFourZero$~m$E_\mathrm h$ in Part I, $\cmpTtcnmFourZero$ instead of $\cmpPartITtcnmFourZero$~m$E_\mathrm h$ for CN$^-$ at 4~\AA, and $\cmpTtbeoFourZero$ instead of $\cmpPartITtbeoFourZero$~m$E_\mathrm h$ for BeO$^+$ at 4~\AA. For N$_2^+$ at 4~\AA{} and for the separated atoms of N$_2^+$ and CO$^+$ the errors, $\cmpTtnTwoFourZero$, $\cmpTtnTwoInf$, and $\cmpTtcoInf$~m$E_\mathrm h$, reproduce those of Part I, and for NO$^+$ and CN$^-$ at 6~\AA{} and the separated atoms of NO$^+$ both agree with FCI within the solver accuracy.

% generated by calc/make_tex.py from calc/derived; do not edit
\begin{table*}[tbp]
\caption{Polyatomic dissociation (STO-6G, $P,Q,G$): errors $E-E_{\mathrm{FCI}}$ in m$E_\mathrm{h}$ after separation
with different cut families (number of cuts in parentheses). Atoms: each atom cell against the rest
(Part I-type regions). Half spaces: planes at the bond midpoints of the chain. Subsets: every union of atom cells
against the rest (two cells). Joint: the occupation law of all atom cells jointly with spin.
$^a$SDPA stops with phase pFEAS and $|E_P-E_D|\approx0.02$~m$E_\mathrm h$ for all H$_2$O solutions
(supplementary material), so the differences between its columns are not significant.}
\label{tab:poly}
\begin{ruledtabular}
\begin{tabular}{lrrrrr}
System & $P,Q,G$ & Atoms & Half spaces & Subsets & Joint\\
\hline
H$_6^+$, 2.0~\AA & $-54.33$ & $-54.33$ (0) & $-36.55$ (4) & $-35.66$ (9) & $-35.66$ (8)\\
H$_6^+$, 2.5~\AA & $-47.71$ & $-47.71$ (0) & $-28.30$ (4) & $-21.41$ (9) & $-21.41$ (10)\\
H$_4^+$, 2.0~\AA & $-6.87$ & $-6.87$ (0) & $-1.96$ (1) & $-1.96$ (2) & $-1.96$ (1)\\
H$_2$O, 2.5~\AA$^a$ & $-1.83$ & $-1.83$ (1) & -- & $-1.83$ (1) & $-1.83$ (1)\\
\end{tabular}
\end{ruledtabular}
\end{table*}

\begin{figure*}[tbp]
 \centering
 \includegraphics[width=\textwidth]{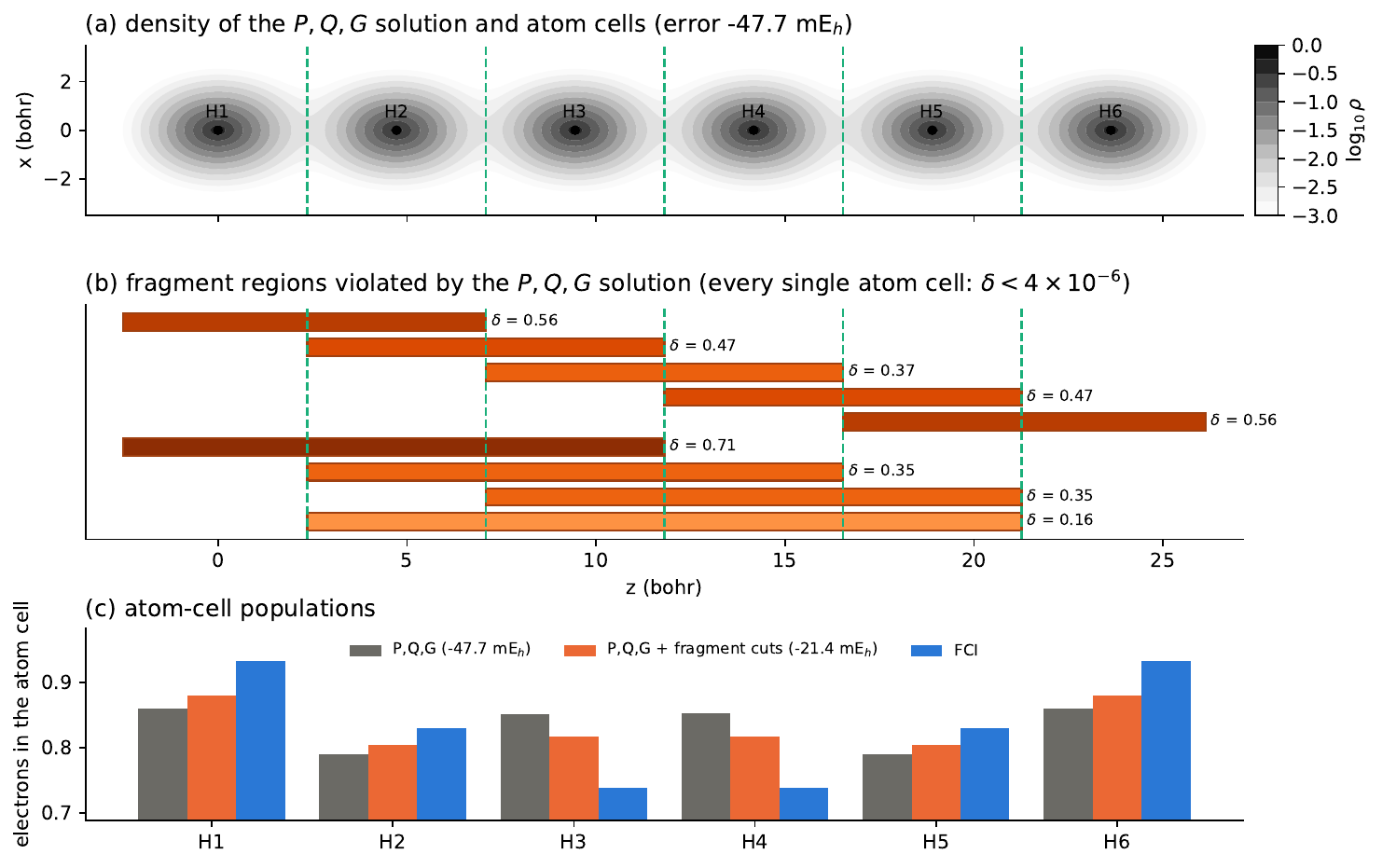}
 \caption{Fragment regions of the H$_6^+$ chain at 2.5~\AA{} (STO-6G, $P,Q,G$). (a) Electron density of the $P,Q,G$ solution without cuts and the atom cells, bounded at the bond midpoints (dashed). (b) Unions of atom cells on which the two-cell LP of this solution is violated, with the LP optimum $\delta$ (one region of each complementary pair); no single atom cell is violated. (c) Electrons in each atom cell for the solution without cuts, the solution with the cuts of all unions of atom cells, and FCI, with the energy errors.}
 \label{fig:fragments}
\end{figure*}

The polyatomic chains distinguish the choice of region from the use of joint laws (Table~\ref{tab:poly}). For H$_6^+$ the relaxed pair density satisfies the two-cell conditions for every single atom cell, so cuts on atomic regions leave the $P,Q,G$ error of $\polyAtomsHsixpTwoFive$~m$E_\mathrm h$ at $2.5$~\AA{} unchanged. Its violations are carried by fragment regions: pairs of neighboring atoms, halves of the chain, and interior triples (Fig.~\ref{fig:fragments}). Subsystem constraints on atom-centered orbital subspaces of F$_3^-$ showed a similar need for constraints on the separated units, including a diatomic one, with an active set that depends on the geometry;\cite{VanAggelen2011} here the regions are real-space regions found by separation, without fragment energies. Planes at the bond midpoints isolate only the end fragments and reduce the error to $\polyHalfspaceHsixpTwoFive$~m$E_\mathrm h$; the two-cell conditions for all unions of atom cells, generated by separation, reduce it to $\polySubsetsHsixpTwoFive$~m$E_\mathrm h$. The joint occupation law of all six atom cells with spin gives the same value, $\polyJointHsixpTwoFive$~m$E_\mathrm h$, and the same holds for H$_6^+$ at $2.0$~\AA{} and for H$_4^+$, so for these chains the energy gain comes from the regions found by separation; the joint laws yield no additional improvement within the reported accuracy, which does not imply that the two feasible sets coincide. For H$_2$O, whose $P,Q,G$ error is small, the cuts have no effect beyond the solver accuracy of that system.

For two cells and spin-summed counts, the count cuts of Part I and the occupation LP are equivalent; Part I also imposed the up-spin and down-spin counts separately, and the joint spin-resolved law added here constrains them together. In the hydrogen chains and, by the two-cell control, in NO$^+$ the energy gains come from two-cell regions chosen by the LP optimum; the LP adds joint multi-cell and spin-resolved conditions, completeness along a refinement, a distance bound, and a constructive repair (a comparison is given in the supplementary material).

The complete problem is hard: on the exposed face $\sum_aD_{aa}=0$ of the occupation polytope every cell holds at most one electron, so linear optimization over the polytope contains the ground-state problem of a classical lattice gas at fixed particle number, an Ising spin glass, which is NP-hard in general (supplementary material).\cite{DezaLaurent1997,Barahona1982} In the systems tested, however, the effective conditions were found among two-cell LPs, which have $N+1$ configurations for spin-summed counts and $(N_\uparrow+1)(N_\downarrow+1)$ for the joint spin-resolved law used in the plane search, together with a one-dimensional search, and the iterations converged in a few steps. Whether this persists for larger molecules is not established; the computational parts are separated by cost in the supplementary material.

\subsection{Fractional electron numbers}
\label{sec:fractional}
Van Aggelen \emph{et al.} traced the fractional charges at dissociation to the energy of the isolated atoms at fractional electron number, which the $P$, $Q$, and $G$ conditions place below the exact piecewise-linear curve.\cite{VanAggelen2009} For $N=k+\theta$ with an integer $k$ and $0<\theta<1$, the exact energy is the lower convex envelope of the integer energies $E_j$; if the chord between $E_k$ and $E_{k+1}$ lies on or below every $E_j$, which holds for the FCI energies of all atoms and intervals considered below with $j$ ranging over all electron numbers from $0$ to $2n$ (supplementary material), it equals $(1-\theta)E_k+\theta E_{k+1}$ and is attained by the ensemble
\begin{equation}
 \rho=(1-\theta)\rho_k+\theta\rho_{k+1}
 \label{eq:ppbl}
\end{equation}
of $k$- and $(k+1)$-electron ground states,\cite{Perdew1982,Yang2000} whose electron count has the variance $\theta(1-\theta)$, the smallest variance of an integer-valued count with mean $k+\theta$. We compare the relaxations of Table~\ref{tab:fractional}; all impose $P$, $Q$, and $G$, and optionally $T_1$ and $T_2$, on Fock-space ensembles that commute with $\widehat N_\alpha$ and $\widehat N_\beta$, which is no restriction because the Hamiltonian conserves both counts. Unlike the molecular calculations, the atomic relaxations impose no spin condition such as $\langle\hat S_-\hat S_+\rangle=0$, neither at the integer endpoints (the fixed-$N$ relaxations) nor at fractional $N$; the FCI references are the lowest states in the prescribed $S_z$ sectors. For an atom, every ground multiplet has a component with the smallest $|M_S|$, so with $|N_\uparrow-N_\downarrow|\le1$ these are the ground states. The derivations are given in the supplementary material.

\begin{table*}[tbp]
\caption{Relaxations of the energy at fractional electron number $N=k+\theta$. At the integer endpoints the fixed-$N$ relaxation imposes both spin traces and the contraction for each spin separately.}
\label{tab:fractional}
\begin{ruledtabular}
\begin{tabular}{p{0.19\textwidth}p{0.39\textwidth}p{0.14\textwidth}p{0.2\textwidth}}
Relaxation & Conditions besides $P$, $Q$, $G$ & Variance of $N$ & Feasible set\\
\hline
Fixed $N$ at fractional $N$ (the form inferred for Ref.~\citenum{VanAggelen2009}) & both spin traces; spin-summed contraction, Eq.~\eqref{eq:contraction}, with $N=k+\theta$ & $0$ & no physical state\\[3pt]
Fock space without cuts (control) & traces of $\gamma_\alpha$, $\gamma_\beta$; $G$ bordered by the identity, Eq.~\eqref{eq:borderedG} & $\ge0$ & every ensemble with the given mean counts\\[3pt]
Fock space with count cuts & in addition Eq.~\eqref{eq:fockcuts} & $\ge\theta(1-\theta)$ & every ensemble with the given mean counts\\[3pt]
Two-sector contractions & in addition Eqs.~\eqref{eq:sectorcontraction} and \eqref{eq:sectorbounds} & $\theta(1-\theta)$ & every ensemble of the form of Eq.~\eqref{eq:ppbl} with prescribed $(N_\uparrow,N_\downarrow)$ at each endpoint\\
\end{tabular}
\end{ruledtabular}
\end{table*}

\emph{Fixed-$N$ conditions at fractional $N$.} Imposing Eq.~\eqref{eq:contraction} with $N=k+\theta$ is, as far as we can infer, the form used in Ref.~\citenum{VanAggelen2009}, whose equations are not stated there. Its trace gives $2\Tr\Gamma=N(N-1)$ and therefore zero count variance, so its feasible set contains no physical state, and its energy is not guaranteed to be either an upper or a lower bound on $E(N)$. The zero variance is a consequence of the trace of Eq.~\eqref{eq:contraction}; Ref.~\citenum{VanAggelen2009} reports fractional occupations and energies, not the count variance. Verstichel \emph{et al.} already called this continuous-$N$ extension unphysical, because it makes no reference to the ensemble, and noted that an SDP on a union of isolated subsystems behaves in the same way for the electron number of each subsystem.\cite{Verstichel2010} The Fock-space relaxation with count cuts contains every physical ensemble with the prescribed mean counts, so its optimum is a lower bound on $E(N)$ and its deviation from the piecewise-linear curve measures what the imposed conditions miss for physical states. The two-sector relaxation contains every two-sector ensemble, so its optimum is a lower bound whenever the exact minimizer is such an ensemble (see below).

\emph{Fock space with integer-count cuts.} Only the traces of $\gamma_\alpha$ and $\gamma_\beta$ are fixed, since Eq.~\eqref{eq:contraction} holds only for a definite electron number. The $G$ condition is bordered by the identity,
\begin{equation}
 \Lcal\bigl((c+A)^\dagger(c+A)\bigr)\ge0,\qquad c\in\mathbb C,
 \label{eq:borderedG}
\end{equation}
for every one-body operator $A$, which is equivalent to Eq.~\eqref{eq:centered}; it follows from $G$ for a fixed electron number but is an additional condition in Fock space. The integer-count inequalities
\begin{equation}
 \Lcal\bigl((\widehat N-j)(\widehat N-j-1)\bigr)\ge0,\qquad j=0,1,\dots,
 \label{eq:fockcuts}
\end{equation}
with $\Lcal(\widehat N(\widehat N-1))=2\Tr\Gamma$, are Eq.~\eqref{eq:integercut} for the single cell $\R^3$ applied to a count that is not fixed; the cut with $j=k$ gives a variance of at least $\theta(1-\theta)$. With the border, zero variance at integer $N$ implies Eq.~\eqref{eq:contraction}; without it, the Fock-space relaxation can lie far below the fixed-$N$ one.

\emph{Two-sector contractions.} For the ensemble of Eq.~\eqref{eq:ppbl}, with sector spin counts $N_s^{(k)}$ and $N_s^{(k+1)}$, let $\gamma_0=(1-\theta)\gamma^{(k)}$ and $\gamma_1=\theta\gamma^{(k+1)}$ be the weighted sector 1-RDMs, so that $\gamma=\gamma_0+\gamma_1$. Each sector satisfies its own contraction, and the weighted sum gives
\begin{subequations}\label{eq:sectorcontraction}
\begin{align}
 2\sum_{q\in s}\Gamma_{pq;rq}={}&(N_s^{(k)}-\delta_{s,\sigma_p})(\gamma_0)_{pr}\\
 &+(N_s^{(k+1)}-\delta_{s,\sigma_p})(\gamma_1)_{pr},
\end{align}
\end{subequations}
where $\sigma_p$ is the spin of orbital $p$, together with
\begin{subequations}\label{eq:sectorbounds}
\begin{align}
 &0\preceq\gamma_0\preceq(1-\theta)I,\qquad
 0\preceq\gamma_1\preceq\theta I,\\
 &\Tr\gamma_{0,s}=(1-\theta)N_s^{(k)},\qquad
 \Tr\gamma_{1,s}=\theta N_s^{(k+1)}.
\end{align}
\end{subequations}
Only the contraction of the total $\Gamma$ and the sector 1-RDMs enter, so $\gamma_0$ is the only additional variable. These conditions fix the variance at $\theta(1-\theta)$ and reduce at $\theta=0$ to the fixed-$N$ relaxation used at the integer endpoints, which imposes both spin traces and the contraction for each spin separately, $2\sum_{q\in s}\Gamma_{pq;rq}=(N_s-\delta_{s,\sigma_p})\gamma_{pr}$. That relaxation implies the spin-summed contraction of Eq.~\eqref{eq:contraction} but is not implied by it: an equal mixture of $|a\!\uparrow b\!\uparrow\rangle$ and $|a\!\downarrow b\!\downarrow\rangle$ satisfies Eq.~\eqref{eq:contraction} with $N=2$ and both spin means equal to one, but has a nonzero same-spin contraction. The fixed-$N$ form at fractional $N$ imposes both spin traces and the spin-summed contraction, Eq.~\eqref{eq:contraction} with $N=k+\theta$. Equations~\eqref{eq:sectorcontraction} and \eqref{eq:sectorbounds} hold whenever the exact minimizer is a two-sector ensemble with these endpoint spin counts, which is the case for a fixed mean count when $E_j$ is convex in $j$; if both spin means are prescribed, the minimizer can combine three spin sectors. The conditions are necessary, not sufficient: they do not ensure that $\Gamma$ splits into sector 2-RDMs that each satisfy $P$, $Q$, and $G$. Requiring that split for the two adjacent sectors gives exactly the straight line between the fixed-$N$ relaxations at $k$ and $k+1$. The fractional-$N$ ensemble representability of Verstichel \emph{et al.} optimizes over all sectors and gives the lower convex envelope of the integer SDP energies, which equals that line when the line supports the envelope; convexity of the FCI energies does not imply that of the SDP energies. They evaluate it through the fixed-$N$ SDP energies computed in advance and use fragment energies in their subsystem constraints.\cite{Verstichel2010} The two-sector contractions are the part of that definition that involves only the sector 1-RDMs, imposed directly in a single SDP without precomputed energies, and form a weaker relaxation. With subsystem constraints the fragment energies become size-consistent, but the 2-RDM of the dissociated system still violates the fragment contractions.\cite{VanAggelen2011}

\begin{figure*}[tbp]
 \centering
 \includegraphics[width=\textwidth]{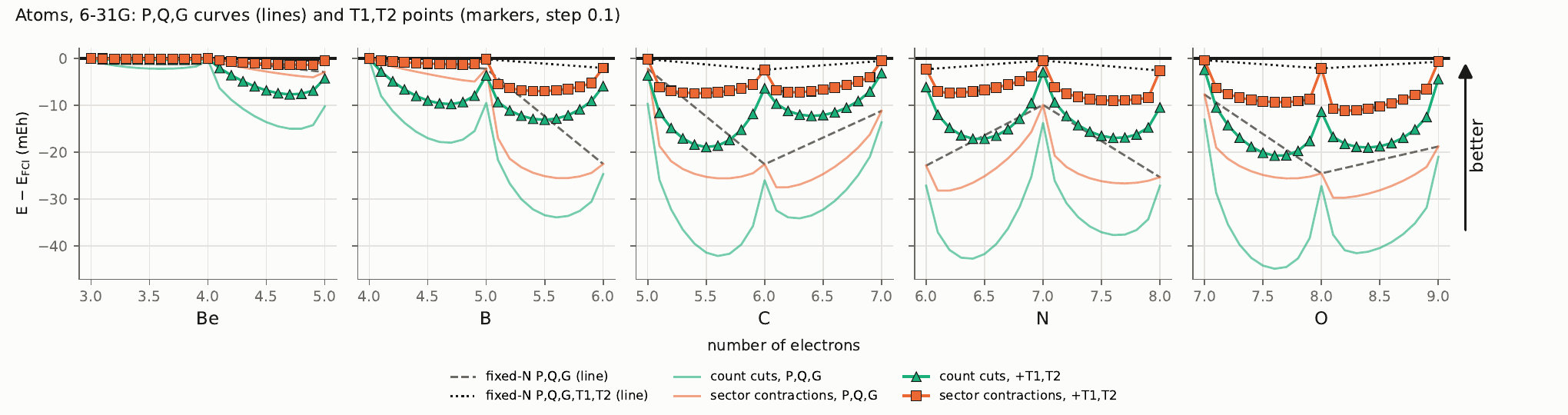}
 \caption{Deviation from the exact piecewise-linear $E(N)$ (FCI at integer $N$) for the atoms Be, B, C, N, and O in the 6-31G basis (canonically orthonormalized atomic orbitals, all electrons). Dashed and dotted lines: straight lines between the fixed-$N$ $P,Q,G$ and $P,Q,G,T_1,T_2$ relaxations at integer $N$. Solid lines: Fock-space relaxation with the bordered $G$ condition and the integer-count cuts (count cuts) and two-sector relaxation (sector contractions), both with $P$, $Q$, and $G$, in steps of 0.1. Markers: the same relaxations with $T_1$ and $T_2$. The added electron of $N=k+\theta$ is assigned to one spin, so that one spin mean stays integer and $|N_\uparrow-N_\downarrow|\le1$ at every integer $N$; the lowest state of each such $S_z$ sector is the ground state, so the exact ensemble of adjacent integer ground states lies on this spin path. The plotted quantity is the signed deviation from the FCI line; the deviations quoted in the text are measured from the common chord of the fixed-$N$ SDP relaxations at the integers.}
 \label{fig:fractional}
\end{figure*}

Figure~\ref{fig:fractional} compares the Fock-space relaxation with integer-count cuts and the two-sector relaxation for the atoms Be to O in the 6-31G basis, each between $Z-1$ and $Z+1$ electrons, against the exact piecewise-linear curve and the straight lines between the fixed-$N$ relaxations.

We quote the deviation from the common chord of the fixed-$N$ SDP relaxations at the integers, rather than the signed deviation from the FCI line plotted in Fig.~\ref{fig:fractional}; for the count-cut relaxation this deviation also contains the offsets of its own values at the integers, and its curvature relative to its own chord is $\fracBcutOwnMin$--$\fracBcutOwnMax$~m$E_\mathrm h$ (supplementary material). In the 6-31G basis the fixed-$N$ form at fractional $N$ deviates from the straight line between the fixed-$N$ $P,Q,G$ relaxations by up to $\fracZerovarMin$--$\fracZerovarMax$~m$E_\mathrm h$ in the different intervals. The same bordered Fock-space relaxation without the count cuts deviates by $\fracBfockMin$--$\fracBfockMax$~m$E_\mathrm h$, as much as the fixed-$N$ form except in the interval $[3,4]$ of Be, where the change of formulation alone reduces the largest deviation by about two thirds, and its count variance stays far below $\theta(1-\theta)$; adding only the cuts of Eq.~\eqref{eq:fockcuts} reduces the largest deviation to $\fracBcutMin$--$\fracBcutMax$~m$E_\mathrm h$. In every interval the cuts improve the bordered Fock-space control, and in all intervals but that of Be they account for the improvement over the fixed-$N$ form. These are interval maxima, possibly at different $N$, and they include the offsets of each relaxation at the integers from the fixed-$N$ endpoints. The two-sector contractions reduce it further to $\fracBsplitMin$--$\fracBsplitMax$~m$E_\mathrm h$. With $T_1$ and $T_2$ the largest deviations from the corresponding $P,Q,G,T_1,T_2$ line are $\fracBcutTMin$--$\fracBcutTMax$ and $\fracBsplitTMin$--$\fracBsplitTMax$~m$E_\mathrm h$ (supplementary material). At integer $N$ the count-cut relaxation lies at most $\fracEndGapMax$~m$E_\mathrm h$ below the fixed-$N$ one ($\fracEndGapTMax$~m$E_\mathrm h$ with $T_1$ and $T_2$), because the Fock-space relaxation does not impose the spin-resolved contractions at integer $N$; with a fixed integer mean, the bordered $G$ condition already makes every count cut nonnegative and does not determine the sector weights. The integer count alone therefore does not restore the piecewise-linear curve; the sector structure of the 1-RDM removes most of the remaining deviation, and the rest is the part that requires the sector structure of the 2-RDM.

The atomic relaxations also describe the separated-atom limit. For NO$^+$ in 6-31G with frozen 1s cores, the sum of a fixed-$N$-at-fractional-$N$ N atom and O atom, minimized over the charge and spin split, has $\fsZerovarNA$ electrons on N, and the $P,Q,G$ molecule at $R=\infty$ has $\fsZerovarMolNA$ electrons on N, an inferred N-region variance of $\fsZerovarMolVar$, and an energy within $\fsZerovarAbsDiff$~m$E_\mathrm h$ of that sum: the $P,Q,G$ molecule is consistent with dissociation into the zero-variance atoms identified by Van Aggelen \emph{et al.}\cite{VanAggelen2009} With the integer-count cuts on the region of the N atom, the molecule has $\fsBcutMolNA$ electrons on N with the variance $\fsBcutMolVar$, equal to $\theta(1-\theta)$, and an error of $\fsBcutMolErr$~m$E_\mathrm h$; the sum of count-cut atoms has its minimum at $\fsBcutNA$ electrons on N and lies $\fsBcutAbsDiff$~m$E_\mathrm h$ below the molecule, whose contractions and inter-fragment $P$, $Q$, and $G$ blocks constrain the fragments more strongly than two independent atoms. In this basis the count cuts correct the fragment statistics but do not restore integer fragment charges. In the minimal STO-6G basis, where few holes are available, the corresponding sums reproduce the molecular energies of NO$^+$, CO$^+$, and N$_2^+$ at $R=\infty$ within $\fsStoMaxDiff$~m$E_\mathrm h$, and NO$^+$ dissociates into integer charges after the cuts (supplementary material); this case is more favorable than the 6-31G one.

\subsection{What the cuts do not remove}
\label{sec:remaining}
After convergence, the partitions active in each run give $\delta\le10^{-5}$ for its final solution (a later test on a finer partition can still detect a violation, as for NO$^+$ in Sec.~\ref{sec:distancebounds}), but substantial errors remain: for example $\cmpAdaptnTwoFourZero$~m$E_\mathrm h$ for N$_2^+$ at 4~\AA{} with $P,Q,G$ and $\cmpTtnTwoFourZero$~m$E_\mathrm h$ with $T_1$ and $T_2$, $\cmpFixedcoInf$ and $\cmpTtcoInf$~m$E_\mathrm h$ for the separated atoms of CO$^+$, and $\polyJointHsixpTwoFive$~m$E_\mathrm h$ for H$_6^+$ at $2.5$~\AA. A finite pass does not certify representability (Sec.~\ref{sec:definitions}), so these solutions are not shown to have representable pair densities; the result is that the remaining errors are not accompanied by any violation that the tested partitions detect. They are therefore not removed by the tested pair-density conditions added to $P$, $Q$, $G$ (or $P$, $Q$, $G$, $T_1$, $T_2$).

The two failure modes identified in Part I distinguish the two diatomic cases. In NO$^+$ and CN$^-$ the relaxation transfers a fraction of an electron between the fragments, and the cuts remove nearly all of the error: at the separated-atom limit of NO$^+$ the remaining $P,Q,G$ deviation of $\cmpFixednopInf$~m$E_\mathrm h$ is smaller than the $\gapNOpInf$~m$E_\mathrm h$ primal--dual gap, and at 4~\AA{} the error with $T_1$, $T_2$, and the searched planes, $\cmpTtnopFourZero$~m$E_\mathrm h$, agrees with FCI within the solver accuracy. In N$_2^+$ the mean count $13/2$ is already correct, and the cuts restore the variance $1/4$, but a large error remains.

For CO$^+$ the remaining error is of a different kind. At its separated-atom limit, Part I found a nonadditive energy error with the $T_1$ and $T_2$ conditions and the region cut, accompanied by an incorrect local spin of the C$^+$ fragment although its electron count was integer to the reported precision (Sec.~V and Table~IV of Part I). The joint spin-resolved occupation laws of Proposition~\ref{prop:spin}, which contain the single-species cuts of Part I and in addition constrain the opposite-spin blocks, were tested here for the real-space atom cells, half spaces, and balls about each nucleus of the separated atoms (Table~\ref{tab:compare}): they remove the remaining occupation violations but leave the residual essentially unchanged. The residual is consistent with an error in the spin-flip part of the local spin, which the counts do not control (Sec.~\ref{sec:spin}), but the calculations show only that the tested occupation conditions do not remove it.

At the separated-atom limit of N$_2^+$ the cuts restore the count statistics of the fragments, but their spin moments violate the identity $\langle\hat{\bm S}_A^2\rangle=3\langle\hat S_{Az}^2\rangle$ that every state of total spin $1/2$ satisfies. Imposing it as an equality changes the energies by at most $\rankTwoShiftMax$~m$E_\mathrm h$ before and $\rankTwoCutShiftMax$~m$E_\mathrm h$ after the cuts, so this violation is not the main source of the remaining error (supplementary material). Conditions on the full 2-RDM, including its spin-flip and off-diagonal parts, are contained in the complete families of two-body witnesses.\cite{Mazziotti2012,RubinDePrince2026}

\section{Scope: probability representability and quantum compatibility}
\label{sec:scope}
\subsection{What is attained in the continuum limit}
The limit of Theorem~\ref{thm:distance} is probability-measure representability of the spatial pair density. At every finite level the lift in Eq.~\eqref{eq:densitylift} is an absolutely continuous $N$-body density with the correct $\rho$, and an antisymmetric sign obtained by ordering the particle coordinates turns its square root into a normalized fermionic wavefunction without prescribed spin. A vanishing distance therefore certifies that a pair density is an $L^1$ limit of pair densities of wavefunctions at the same $\rho$; that the limit itself need not come from a wavefunction is shown by an example in the supplementary material. No uniform kinetic-energy bound, Coulomb-energy convergence, or fixed finite-basis realization follows from an $L^1$ pair-density estimate; the Coulomb kernel is unbounded, so $L^1$ convergence alone does not control its expectation.

\subsection{The separate 1-RDM consistency problem}
\label{sec:onerdm}
For a nonrelativistic Hamiltonian with local pair interaction, an admissible state permits, in the coefficient convention of Eq.~\eqref{eq:brackets}, an energy expression of the form
\begin{equation}
 E[\gamma,n_2]=\gfun h
       +\frac12\iint w(x,y)n_2(x,y)\dd x\dd y.
 \label{eq:jointenergy}
\end{equation}
This identity does not provide the admissible domain. It requires one common $N$-electron state that realizes both $\gamma$ and $n_2$. Separate representability and matching one-particle diagonals are not a replacement for joint realizability. Boundary formulations of RDM and density-functional theories and the compatibility constraints in implicit density-functional theory motivate this distinction.\cite{Percus2005,LiuPercus2006} The present paper imposes no independently prescribed $\gamma$ in Eq.~\eqref{eq:distance_definition}. It therefore neither solves joint representability nor provides a new exact ground-state energy functional.

If valid spatial cuts are added to a fixed finite-basis RDM optimization while retaining all earlier constraints, the exact variational lower bounds can only increase and cannot exceed the fixed-basis exact energy; Part I gives molecular examples for two-region cuts.\cite{NakataPartI} The present results do not show that their limit equals that energy: even a complete spatial hierarchy leaves the non-diagonal and joint-compatibility problem. Likewise, the constructed pair-density approximants do not necessarily belong to the input Gaussian orbital space. These are restrictions on the claim, not failures of the fixed-$\rho$ pair-density theorem.

\subsection{Relation to prior diagonal and pair-density theories}
The occupation inequalities and the Slater hull,\cite{Kummer1967,WeinholdWilson1967,McRaeDavidson1972,Davidson2003,AyersDavidson2007} the integer-count bound,\cite{Yamada1961,Erdahl1987} its use as a molecular cut,\cite{NakataPartI} measure extension under coarse graining,\cite{KonstantopoulosYuan2019,Carlier2023} particle-number extension hierarchies and spatial-partition tests,\cite{Friesecke2013,Dusson2025} and localized representable constructions\cite{Pistol2004} are known; the present hierarchy is not asserted to be the first systematic one, and a $k$-particle probability extension and a degree-$k$ CAR Gram relaxation are distinct tests. The contribution here is an exact finite occupation distance paired with an explicit $\rho$-preserving lift and a vanishing error gap, with dual certificates and Gaussian examples of what 3-positivity can miss; the five-electron input is taken from Ref.~\citenum{NakataDraft}, with its degree-three positivity proved here. The transfer does not assert that every diagonal violation is detectable in a preassigned molecular basis.

\section{Conclusions}
\label{sec:conclusions}
We have placed the integer-count conditions of Part I as the first members of a convergent hierarchy of rigorous relaxations for the real-space pair density at fixed one-particle density, defined by finite integer occupation linear programs on spatial partitions. Every level contains all representable pair densities and so, added to a 2-RDM relaxation in a finite basis, gives lower bounds to the full configuration-interaction energy in that basis, the levels tighten monotonically under nested refinement, and their intersection is the set of pair densities with a symmetric $N$-particle probability-measure extension; the energy bounds increase monotonically, but their limit can lie strictly below the full configuration-interaction energy, for a nonlocal Hamiltonian and, in a suitably chosen basis, for the Coulomb repulsion (supplementary material); whether a gap remains for the molecules computed here, with atom-centered Gaussian bases, is not decided. Building on known realizability and exchangeability results, we have established three properties of this hierarchy. First, the spin-summed real-space count conditions of Part I are exactly its two-cell members (Part I also used single-spin and orbital-region cuts); the hierarchy adds joint multi-cell conditions and, at fixed $S_z$, joint spin-resolved conditions that detect violations missed by separate charge and single-spin counts, and it is complete under refinement. Second, the LP optimum is a distance and not only a test: $\delta_{\mathcal P}\le d_\rho(n_2)\le\delta_{\mathcal P}+e_{\mathcal P}$, with $\delta_{\mathcal P}$ monotone under refinement and $e_{\mathcal P}\to0$; its dual solution certifies exclusion, and its primal law constructs a representable $N$-body density with the same $\rho$. Third, these conditions are not implied by 3-positivity: moment data that satisfy all three-particle positivity conditions, including $T_1$, $T_2$, and $T_2'$, and have a pure-state-representable 1-RDM give through Gaussian orbitals a smooth, nonnegative pair density with a positive semidefinite density covariance that is incompatible with an integer electron count in a half space, the real-space signature of fractionally charged fragments. A rational benchmark verifies the bracket and the repair exactly.

Applied to variational 2-RDM solutions in the minimal STO-6G basis without any Hartree--Fock input, the LP turns these properties into computations. For N$_2^+$ at 4~\AA{} the $P,Q,G$ pair density lies, by the numerical LP estimate, at least $\ePdeltaPqgnTwoFourZero$ (of an ordered-pair mass of $\ePmassnTwoFourZero$) from every representable pair density with the same $\rho$. The dual witnesses, added as cuts, reproduce the region cuts of Part I without a candidate set where the fragments of Part I carry integer or half-integer charges, and in all tested cases with fractional charge transfer the plane chosen by the LP optimum gives smaller errors: $\cmpAdaptnopTwoFive$, $\cmpAdaptnopThreeZero$, $\cmpAdaptnopThreeFive$, and $\cmpAdaptnopFourZero$~m$E_\mathrm h$ for NO$^+$ at 2.5, 3, 3.5, and 4~\AA, compared with $\cmpPartInopTwoFive$, $\cmpPartInopThreeZero$, $\cmpPartInopThreeFive$, and $\cmpPartInopFourZero$~m$E_\mathrm h$ in Part I, and with $T_1$ and $T_2$ $\cmpTtnopTwoFive$ to $\cmpTtnopFourZero$~m$E_\mathrm h$ at 2.5 to 4~\AA{} ($\cmpPartITtnopTwoFive$ to $\cmpPartITtnopFourZero$~m$E_\mathrm h$ in Part I). For NO$^+$ at 4~\AA{} the $L^1$ distance of the pair density from the FCI one falls from $\fciDistPqgnopFourZero$ to $\fciDistCutsnopFourZero$. In polyatomic chains the LP locates fragment regions that no single plane per bond isolates, and the cuts reduce the error of H$_6^+$ at 2.5~\AA{} from $\polyAtomsHsixpTwoFive$ to $\polySubsetsHsixpTwoFive$~m$E_\mathrm h$.

The count conditions also clarify the energy at fractional electron number. The fixed-$N$ conditions at fractional $N$, the form inferred here for the analysis of fractional dissociation, imply through their trace atoms with zero count variance, which are not physical states. Fock-space relaxations with the bordered particle--hole condition and the integer-count inequalities contain every physical ensemble and give lower bounds on $E(N)$. For the atoms Be to O in the 6-31G basis they reduce the largest deviation from the straight line between the fixed-$N$ relaxations from $\fracZerovarMin$--$\fracZerovarMax$ to $\fracBcutMin$--$\fracBcutMax$~m$E_\mathrm h$, whereas the same Fock-space relaxation without the cuts deviates by $\fracBfockMin$--$\fracBfockMax$~m$E_\mathrm h$, and contractions resolved by electron-number sector reduce it to $\fracBsplitMin$--$\fracBsplitMax$~m$E_\mathrm h$ ($\fracBsplitTMin$--$\fracBsplitTMax$~m$E_\mathrm h$ with $T_1$ and $T_2$). At the separated-atom limit the $P,Q,G$ solution of NO$^+$ agrees with the sum of zero-variance atoms within $\fsZerovarAbsDiff$~m$E_\mathrm h$; the count cuts restore the fragment-count variance $\theta(1-\theta)$, but in 6-31G they do not restore integer fragment charges.

These results also delimit what pair-density conditions can do. The guarantee concerns the pair density only: compatibility with a prescribed 1-RDM, kinetic-energy regularity, and realization in a fixed Gaussian orbital space require further information, which we deliberately do not derive from the pair-density hierarchy. In the molecules tested, errors remain after every tested partition passes, for example $\cmpAdaptnTwoFourZero$~m$E_\mathrm h$ for N$_2^+$ at 4~\AA{} with $P,Q,G$ and $\cmpTtnTwoFourZero$~m$E_\mathrm h$ with $T_1$ and $T_2$, and joint multi-cell laws gave no gain beyond two-cell fragment regions within the reported accuracy: in H$_6^+$ the energy gain came from finding the fragment regions, whereas the strength of the joint conditions is shown by the exact three-cell example of Sec.~\ref{sec:definitions}. The same construction applies to $k$-particle densities, with cell moments of order $k$ and count polynomials of degree $k$; for $k\ge3$ it requires a lifted $k$-RDM in a variational calculation. Efficient separation over many cells, larger basis sets, and conditions beyond the pair density remain open.

\section*{Supplementary material}
The supplementary material gives independent exact CAR audits of degree-three positivity; Gaussian localization and physical-state controls; the one-hole qualification for $P+Q$; the partition characterization as background; the $L^1$ closure at fixed $\rho$ with the measure-versus-wavefunction example; the proof of the convergence of the coarse-graining error; a conditional higher-order transfer statement; the molecular settings, solver accuracy, and further molecular results; the derivations for fractional electron numbers; the occupation LP in standard form with its dual; and details of the distance certificates. Source code, repaired occupation weights, dual witnesses, and reproduction logs accompany the manuscript bundle.

\section*{Author declarations}
\subsection*{Conflict of interest}
The author has no conflicts to disclose.

\subsection*{Use of AI tools}
The author used Claude Code (Anthropic), an AI coding assistant, in this work: to write and run the verification programs, to search for and check bibliographic data, and to draft parts of the manuscript text. The author also used OpenAI's GPT-6 Astra model, through ChatGPT, mainly to review drafts of the manuscript: its comments on the arguments and the literature were checked by the author against the proofs, the verification output, and the cited sources before any change was made. The author designed the study, directed and checked this work, and takes full responsibility for the content of this article.

\section*{Data availability}
The code, the exact verification programs and their reports, and the data that support the findings of this study are openly available in Zenodo at \url{https://doi.org/10.5281/zenodo.24063922}.\cite{NakataZenodo2026} The molecular calculations are reproduced by the programs in the \texttt{calc} directory of that record, which also generate the tables of Sec.~\ref{sec:molecules}; no external benchmark data are used.

\bibliographystyle{aipnum4-2}
\bibliography{references}
\end{document}